\documentstyle[preprint,aps,prb]{revtex} 
\begin{document}
\draft
%
%
\title{Berry-phase treatment of the homogeneous electric field
perturbation in insulators}

\author{R. W. Nunes$^1$ and Xavier Gonze$^2$}

\address{$^1$Departamento de F\'{\i}sica, C.P. 702, Universidade Federal de Minas Gerais - Av. Ant\^onio Carlos 6627 - Belo Horizonte MG - 30123-970 - Brazil//
$^2$Unit\'e de Physico-Chimie et de Physique des Mat\'eriaux, Universit\'e Catholique de Louvain, B-1348 Louvain-la-Neuve, Belgium}

\date{\today}
\maketitle

\begin{abstract}
A perturbation theory of the static response of insulating
crystals to homogeneous electric fields, that combines the
modern theory of polarization with the variation-perturbation
framework is developed, at unrestricted order of perturbation.
Conceptual issues involved in the definition of such a
perturbative approach are addressed.  
In particular, we argue that the position
operator, ${\hat x}$, can be substituted by the
derivative with respect to the wavevector, $\partial/\partial k$,
in the definition of an electric-field-dependent energy functional 
for periodic systems.
Moreover, due to the unbound nature of the perturbation, a regularization 
of the Berry-phase expression for the polarization is needed in order to 
define a numerically-stable variational
procedure. Regularization is achieved by means of a discretization procedure,
which can be performed either before or after the perturbation
expansion. We compare the two possibilities, show that
they are both valid, and analyze their behavior when applied to a
model tight-binding Hamiltonian.
Lowest-order as well as generic formulas are presented     
for the derivatives of the total energy, the normalization condition,
the eigenequation, and the Lagrange parameters.
\end{abstract}

\narrowtext

\section{Introduction}
\label{sec:intro}

Until the early nineties, the formulation of a quantum-mechanical
theoretical framework for the study of the physics of electric
polarization in solids had remained a challenging
problem.~\cite{footnote1,resta1} Even the definition of the polarization
itself as a bulk quantity, independent of surface termination, was the
subject of heated debate.~\cite{martin1,tagantsev,baldereschi}

This picture changed when King-Smith and Vanderbilt (KS-V)~\cite{ksv}
proposed a formulation (the modern theory of polarization - MTP),
which resolved the conceptual difficulties associated with the
definition of this quantity for continuous, periodic, charge
distributions. In their work, the electric polarization of an
insulating crystal is related to a Berry phase~\cite{berry} computed
from the valence wavefunctions. The existence of a band-structure
Berry phase had already been discussed by Zak and
coworkers,~\cite{zak1} before its connection with the
electronic polarization was established by KS-V. Besides settling the
important conceptual question related to the definition of
polarization as a bulk quantity, the KS-V theory provided an entirely
new framework for the computation of the polarization of a crystal
maintained at vanishing homogeneous electric field.  Since its
formulation, the theory has been examined in greater detail by
KS-V~\cite{vks} and by Resta~\cite{resta1}, and extended to many-body
systems by Ortiz and Martin~\cite{ortiz1}. The relation between
polarization and the phases of the wavefunctions has also led to a
reexamination of the role this quantity plays in the Density
Functional Theory (DFT)~\cite{hk,ks} formulation of the ground-state
properties of extended
systems.~\cite{ggg1,ggg3,ggg2,resta2,dhv,martin2}

Of no less importance is the conceptual relationship between the
spontaneous polarization and the centers of charge of the Wannier
functions (WF) of the occupied bands, which was also discussed by
KS-V~\cite{ksv} and previously by Zak.~\cite{zak1,zak2} This connection was
later generalized by Nunes and Vanderbilt~\cite{nv} (NV) to deal with
insulators placed in non-zero external homogeneous electric fields:
they introduced field-dependent ``polarized'' WF's and a method for
their computation. NV argued that, in the static-response
regime, the state of an insulator under an external homogeneous
electric field is one in which the periodicity of the charge density
is retained, despite the fact that the perturbation lacks the
lattice-translational symmetry of the unperturbed crystal. Such state
is actually a long-lived resonance of the system, as rigorously
demonstrated by Nenciu~\cite{nenciu}.

It is well known that within DFT~\cite{hk,ks}, ground-state properties
of condensed-matter systems, such as equilibrium lattice parameters,
bond lengths and bond angles, among others, can be obtained with an
accuracy of a few percent in comparison with experimental
results. 
Within DFT~\cite{hk,ks}, the NV method has recently 
been applied to the computation of the
polarized WF's and the dielectric constant of silicon and gallium
arsenide~\cite{fernandez}.
The latter quantity is related to the change of polarization due to a
change of homogeneous electric field, in the linear regime,
or equivalently, to the second-derivative of the total energy
with respect to the homogeneous electric field.

Specific treatments have been developed (noticeably within DFT) for
the study of the response of crystals to ``external'' perturbations,
like phonons, stresses or homogeneous electric fields.  The latter, on
which we focus exclusively, can be taken as a homogeneous field or as
the limit of long-wavelength perturbations.

In the so-called direct
approach~\cite{fleszar}, supercell calculations are employed to study
both the unperturbed and perturbed systems, with the response
functions being obtained by numerical finite-difference analysis of
the changes induced by a long-wavelength perturbation applied to
elongated supercells.  The non-linear response
regime is directly accessible, although it must be disentangled from
the linear response of the system.  However, because of the use of
supercells, the computational cost of this approach 
is rather high.

Alternatively, the specific response to a homogeneous electric field
was considered within perturbation theory, already in the sixties.  In
the Random-Phase Approximation~\cite{adler} (no local-field
effects~\cite{footnote3} included), the response of the wavefunctions
is obtained through a sum over states, involving matrix elements of
the position operator between valence and conduction
states~\cite{baroni1}. This technique was generalized to the
computation of second- and third-order
susceptibilities~\cite{higher_indpt}.  The need to compute many
unoccupied bands is the bottleneck of this method.

Local-field effects can be reintroduced on top of such a sum-over-states
approach either in a matrix-inversion framework~\cite{wiser,hybers},
or in an iterative approach~\cite{levine_allan}. 
In their calculation of linear susceptibilities, Levine and Allan
included a so-called ``scissor-operator'' correction, that was
understood later to compensate some deficiencies of 
local-density approximation computation of
long-wavelength response functions~\cite{ggg1}. Also, 
Levine and coworkers proposed rather
involved expressions for the second-order and
third-order DFT susceptibilities~\cite{higher_localfields}. 

To a large extent, Density Functional Perturbation Theory
(DFPT)~\cite{baroni2,giannozzi,zein,gonze1,gonze2,gonze6,gonze3,gonze4}
overcome the limitations of the previously mentioned approaches, at
the price of non-negligible additional coding.  At the lowest order in
the homogeneous-electric-field perturbation, this method was
introduced by Baroni and collaborators~\cite{baroni2,giannozzi}.  It
is based on an iterative solution for the first-order change in the
wavefunctions, which allows for the self-consistent inclusion of
local-field effects, besides eliminating the cumbersome sum over
conduction bands. It does not employ supercells, and can be applied to
perturbations of arbitrary wavelengths. In the DFPT, the computational
workload involved in the computation of linear-response functions is
of the same order of that involved in one ground-state calculation.

DFPT is part of a class of formalisms in which perturbation
theory is applied to a variational principle~\cite{gonze5}.
This interesting combination leads to a generic 
``2n+1'' theorem~\cite{gonze1,gonze6}, as well as variational 
properties of even-order derivatives of the energy~\cite{gonze2,gonze6}.
For example, one can compute the third-order derivative of the energy from
the first-order derivative of wavefunctions, and the fourth-order
derivative of the energy is variational with respect to 
the second-order derivative of wavefunctions.
The expressions derived in this framework are surprisingly simple
and can be formulated at all order of perturbations.

However, the treatment of homogeneous electric fields in this
variation-perturbation framework is plagued by
difficulties similar to those encountered in the theory of
polarization.  Shortly after the appearance of the MTP, Dal Corso
and Mauri~\cite{dalcorso1}, building upon the NV work, proposed a very
concise expression for the second-order susceptibility which was later
applied successfully to compute this quantity for a variety of
systems~\cite{dalcorso2}.

In the present work, we formulate a perturbation theory of the static
response of insulating crystals to homogeneous electric fields that
combines the conceptual ideas of the MTP
with the variation-perturbation theory.  A major achievement of our
work is the presentation of formulas valid for unrestricted order of
perturbation theory.  We also examine the low-order expressions in
some detail, and recover the expression proposed by Dal Corso and
Mauri~\cite{dalcorso1}.

The theory is worked out directly in reciprocal space, in terms of the
Berry phase associated with the occupied bands of the perturbed
crystal, in the manner of the MTP. The
conceptual issues involved in the definition of a perturbative
approach for the problem are addressed. The Berry phase is argued to
remain a valid concept in the presence of the periodicity-breaking
electric field. The periodicity of the charge density is assumed to
survive the application of the field, and the Berry phase is obtained
from the associated polarized periodic wavefunctions. By working out
the perturbative approach in terms of these polarized states, we
obtain very compact expressions for the high-order dielectric-response
functions of the crystal. These can be numerically obtained on the
basis of iterative equations for the second- and higher-order terms in
the perturbation expansion of the wavefunctions of the system, as in
the DFPT approach for other types of perturbation. We will not deal
explicitly with the exchange and correlation parts of the DFT
functional: the main difficulties that we want to address in this
paper are not related to them. The formalism can be extended to
include exchange and correlation terms in a self-consistent fashion,
in the manner presented in Refs~\onlinecite{baroni2,gonze3}.

Any application of the MTP involves a
discretization of the Berry-phase expression, in terms of a series of
wavevectors for the electronic wavefunctions. We have discovered that
such a discretization, that appears naturally also in the present
framework, can be performed at two different conceptual levels when
merged with perturbation theory: either after the derivation of formal
expressions at different orders of perturbation theory, starting from
a continuous Hamiltonian, or at the level of the field-dependent
Hamiltonian itself, before any perturbation expansion is performed.
We will refer to the first approach as the discretization after
perturbation expansion (DAPE) formulation, and to the second as the
perturbation expansion after discretization (PEAD) formulation.

In order to judge the relative merits of these two approaches, 
and also the correctness of the global framework, we analyze
the behavior of a model tight-binding Hamiltonian, for which
analytical responses to an electric-field perturbation have been
obtained up to the fourth order. 

The paper is organized as follows. In the next section, 
we address the conceptual issues
associated with the definition of a Hamiltonian and its perturbative
expansion for the electric-field problem.
Sec.~\ref{sec:VPT} summarizes
the main results of the variational perturbation theory which are used
in this work. In Sec.~\ref{sec:pt-cont},
we work out the continuous formulation of the problem
and its perturbation expansion, from which we
obtain the DAPE version of our theory. In
Sec.~\ref{sec:pt-P}, we work from the start using a discretized
expression for the polarization, which leads to the alternative
PEAD formulation. The theory is applied to a model
one-dimensional system in Sec.~\ref{sec:1d-model}.


\section{Insulators in an electric field: conceptual considerations}
\label{sec:efp}


\subsection{The modern theory of polarization}
\label{sec:mtp}

In the MTP~\cite{ksv,vks,resta1}, the
change in electric polarization per unit volume induced by an
adiabatic change in the crystalline potential (the self-consistent
Kohn-Sham potential in the context of DFT) is written
\begin{eqnarray}
\label{Delta-P}
{\bf \Delta P} = \int_{\lambda_1}^{\lambda_2}\frac{\partial {\bf P}}{\partial\lambda}
~\! d\lambda = {\bf P}(\lambda_2) - {\bf P}(\lambda_1)\,,
\end{eqnarray}
with ${\bf P}(\lambda)$ given in terms of a Berry phase
associated with the occupied bands of the system
\begin{eqnarray}
\label{P-bphase}
{\bf P}(\lambda) = -\frac{2ie}{(2\pi)^3}\sum_n \int_{BZ} d{\bf k}~\!
\left\langle u_{n{\bf k}}^{(\lambda)} \left\vert 
\vphantom{u_N^A}
~\! {\bf \nabla_k} ~\! \right\vert u_{n{\bf k}}^{(\lambda)} \right\rangle \,,
\end{eqnarray}
where $-e$ is the electron charge, $\lambda$ is a parameter
representing the adiabatic change in the potential, and the factor of
2 in the numerator accounts for spin (in this work we will consider
only spin-unpolarized systems).

The gauge relation between periodic functions $u_{n{\bf k+G}}({\bf r})
= e^{-i{\bf G\cdot r}}u_{n{\bf k}}({\bf r})$ is established by
requiring that the Bloch eigenstates be periodic in reciprocal space,
i.e. $\psi_{n{\bf k}} = \psi_{n{\bf k}+{\bf G}}$, where ${\bf G}$ is a
reciprocal-lattice vector. With this choice of gauge, the polarization
changes given by Eq.~\ref{Delta-P} are defined to within a factor of
$(2 e / \Omega) {\bf R}$, where ${\bf R}$ is a lattice
vector.~\cite{ksv} Eq.~\ref{P-bphase} was derived under the
restriction that the macroscopic electric field inside the crystal
vanishes. Moreover, it also requires that the set
of wavefunctions be differentiable with respect to {\bf k}.~\cite{blount}

The actual evaluation of the polarization in Eq.~\ref{P-bphase} is
carried out on a discrete mesh of points in reciprocal space. Because
this expression depends on the phase relationships of wavefunctions at
different {\bf k}-points, the following discretized version was
proposed by KS-V:
\begin{eqnarray}
\label{P-discr}
P_{\parallel}(\lambda) = \frac{e}{4\pi^3}\int d{\bf k}_\perp
\sum_{j = 1}^{N_k} {\rm Im}\left\{ \ln \det \left[
\left. \left\langle ~\!u_{n{\bf k}_j}^{(\lambda)}~\!\right|~\!u_{m{\bf
k}_{j+1}}^{(\lambda)}~\! \right\rangle\right]\right\}\;;
\end{eqnarray} 
where $P_{\parallel}$ is the component of the polarization along the
direction of a short reciprocal-lattice vector, ${\bf G}_\parallel$,
and $N_k$ is the number of {\bf k}-points sampling the Brillouin zone
along that direction for each value of ${\bf k}_\perp$, with ${\bf
k}_j = {\bf k}_\perp + j{\bf G}_\parallel/N_k$.

From a calculational point of view, this discretized expression
ensures that the final result is unaffected by random numerical phases
which may be introduced in the wave functions at different {\bf
k}-points, when these are independently determined by the
diagonalization routine. However, Resta has taken the view that the
discretized expression is to be regarded as more fundamental than the
continuous form.~\cite{resta3} For the formulation of the
electric-field response that we develop in the present work,
discretization is crucial in order to define a numerically-stable
minimization procedure. We will come back to this point in
Sec.~\ref{cutoff}.

The Berry-phase expression can be transformed into a real-space
integral involving the Wannier functions of the occupied bands,
leading to a physically-transparent expression for the polarization in
terms of the centers-of-charge of the Wannier
functions:~\cite{ksv,zak1,zak2,vks,blount}
\begin{eqnarray}
\label{P-wannier}
{\bf P}(\lambda) = -\frac{2e}{\Omega}~\!\sum_n \int {\bf r}~\!
\left|~\! w_n^{(\lambda)}~\! \right|^2 d{\bf r}\,,
\end{eqnarray}
where $\Omega$ is the unit-cell volume.

In principle, the above expressions are valid only at vanishing
electric field. However, it was soon realized~\cite{nv,dalcorso1} that
Eq.~\ref{P-wannier} could be extended to the non-zero field problem,
by introducing the so-called polarized Wannier functions. Polarization
effects were then related to the field-induced shifts in the
Wannier-function centers of charge. In the present work,
Eq.~\ref{P-bphase} is argued to apply to the non-zero-field problem,
thus defining a field-dependent Berry phase containing the
polarization effects. This allows us to work out a perturbative
approach for the finite-field problem.


\subsection{Definition of the energy}

The study of the problem of insulators under an external electric
field has traditionally met with conceptual difficulties, related to
the non-analyticity of the perturbation, as discussed in detail by
Nenciu~\cite{nenciu}. Upon the application of the external field, the
spectrum of electronic states changes non-analytically, with the band
structure of the insulator at zero field being replaced by a continuum
of eigenvalues spanning the entire energy axis (from $-\infty $ to
$+\infty $), even for a field of infinitesimal strength. From a
mathematical point of view, the unbound nature of the perturbation
term, $e~\!{\bf \cal E}{\bf \cdot r}$ (hereafter, we use $\cal E$ for
the magnitude of the electric field), hinders the straightforward
application of perturbation theory, since the diagonal elements of the
position operator in the basis of the unperturbed Bloch states are
ill-defined~\cite{martin2,nenciu,resta4}. Strictly speaking, an
infinite crystal in the presence of an external electric field does
not have a ground state~\cite{ggg3,nv}.

From the physical point of view, in the limit of weak to moderate
fields, the tunneling currents (which destroy the insulating state at
sufficiently high fields) can be neglected, and only the polarization
of the electronic states by the external field is considered. 
One is left with a picture of the problem in which the insulating
state of the unperturbed crystal is preserved, hence the band
structure and the periodicity of the charge density are
retained. The problem is thus physically well defined. The theory we
develop below will concern this periodic polarized insulating state,
resulting from the application of the electric field. Avron and
Zak~\cite{avron} have discussed the stability of the band structure in
the present context, while, as mentioned in the introduction, 
Nenciu~\cite{nenciu} has rigorously shown
such state to be a long-lived resonance of the problem, in the regime
of weak fields.

Building on the conceptual framework set by the MTP, NV introduced a
practical real-space method to handle the problem on the basis of the
WF formulation of the polarization. They noted that this polarized insulating
state can be represented in real space by a set of
field-dependent polarized WF's. In this way, the relationship between
the WF's centers of charge and polarization can be extended to the
non-zero field situation, and an energy functional is defined as
follows
\begin{equation}
\label{E-nv}
E\left [\{w^{\cal E}\},\bf {\cal E}\right ] = E^{(0)}\left [\{w^{\cal E}\}\right ]
-\Omega\,{\bf {\cal E} \cdot}{\bf P}\left [\{w^{\cal E}\}\right ]\;,
\end{equation}
where $\{w^{\cal E}\}$ is the set of field-dependent Wannier functions.

Because the state underlying the above expression is not a
ground-state, rather a resonance, the energy functional is only well
defined for WF's of finite range.~\cite{nv} The truncation of the WF's
provides a mathematical procedure for the regularization of the
problem.  We will come back to this pathology in section~\ref{cutoff}

As a corollary of the existence of polarized WF's, we consider now the
representation of the system in terms of polarized Bloch orbitals.  In
the following sections, we develop two alternative formulations in which
the $k$-space MTP expressions for the electronic polarization,
Eqs.~\ref{P-bphase} and \ref{P-discr}, are extended to the
non-zero-field problem. We will also have to
regularize the ensuing expressions, which is possible by means of
discretization of the $k$-space integrals in such a way that
polarized valence bands are stable against mixing with the conduction
bands.

For simplicity, we concentrate on a one-dimensional non-interacting
spin-unpolarized system. The Hamiltonian for the unperturbed periodic
insulator is given by
\begin{eqnarray}
H^{(0)} = K + V_0\;,
\end{eqnarray}
where $K$ is the kinetic-energy operator, and $V_0$ is a periodic
potential, i.e., $\left[V_0,T_\ell\right] = 0$, where $T_\ell$ denotes a
translation by a lattice vector.  The zero-field function
$u_{nk}^{(0)}$ is the periodic part of the unperturbed Bloch-orbital,
obeying the eigenvalue equation $H^{(0)}_k \left. \left| u^{(0)}_{nk}
\right.\right\rangle = \varepsilon^{(0)}_{nk}\left. \left| u^{(0)}_{nk} \right.\right\rangle $, where
\begin{eqnarray}
H^{(0)}_k = e^{-i k{\hat x}}H^{(0)}e^{i k{\hat x}}
\end{eqnarray}
is the unperturbed cell-periodic Hamiltonian.  

Under the action of an external electric field, the Hamiltonian
becomes
\begin{eqnarray}
H = H^{(0)} + e{\cal E} {\hat x} \;.
\end{eqnarray}
As discussed above, this Hamiltonian is not amenable to a conventional
perturbation treatment. We can arrive at an expression that is
applicable to extended periodic systems, from the following
considerations. 

First, let us assume that we can define a set of field-dependent
cell-periodic functions, representing the polarized state of the
system. These are the Fourier transform of the field-dependent
Wannier functions introduced by NV. Eq.~\ref{P-bphase} is thus
extended to the non-zero field problem, with
\begin{eqnarray}
\label{P-ef}
P\left({\cal E}\right) = -\frac{ie}{\pi}\sum_n \int_0^{\frac{2\pi}{a}}
dk~\!\left\langle u^{\cal E}_{nk}\left|~\frac{\partial}{\partial k}
~\right|u^{\cal E}_{nk}\right\rangle\;.
\end{eqnarray}
defining a field-dependent polarization, including the spontaneous and
induced parts of this quantity.

Next, combining this definition with Eq.~\ref{E-nv} we write
\begin{eqnarray}
\label{E-def}
E = E^{(0)} - a P\left({\cal E}\right){\cal E} = \frac{a}{\pi}
\int_{0}^{\frac{2\pi}{a}}dk \sum_{n=1}^N \left[
\left\langle u^{\cal E}_{nk}\left|H^{(0)}_k \right| u^{\cal E}_{nk}\right\rangle\ + 
{\cal E}\left\langle u^{\cal E}_{nk}\left| 
\left(ie\frac{\partial}{\partial k}\right) 
\right|u_{nk}^{\cal E}\right\rangle\right]\;,
\end{eqnarray}
for the total energy in terms of field-dependent cell-periodic
functions, with the unit-cell volume $\Omega = a$ for our 1D system.

Consider now the expansion of the set $\{u^{\cal E}_{nk}\}$ in terms of
the complete set of zero-field periodic functions:
\begin{eqnarray}
\label{u-expand}
\left.\left| u^{\cal E}_{nk}\right\rangle \right. = \sum_{m=1}^\infty 
C^{\cal E}_{nm}(k) \left.\left| u^{(0)}_{mk}\right\rangle\right. \;.
\end{eqnarray}
In terms of this expansion, Eq.~\ref{E-def} is written
\begin{eqnarray}
\label{E-def-expand}
E = \frac{a}{\pi}
\int_{0}^{\frac{2\pi}{a}}dk \sum_{n=1}^N \sum_{m,m^\prime =1}^\infty 
C^{\cal E\ast}_{nm}(k) C^{\cal E}_{nm^\prime}(k) \left[ 
\left\langle u^{(0)}_{mk}\left|H^{(0)}_k \right| u^{(0)}_{m^\prime k}\right\rangle\ + 
{\cal E}\left\langle u^{(0)}_{mk}\left| 
\left(ie\frac{\partial}{\partial k}\right) 
\right|u^{(0)}_{m^\prime k}\right\rangle\right]\;.
\end{eqnarray}

This expression is in the form of the matrix representation, in the
basis $\{u^{(0)}_{mk}\}$, of the expectation value of the
following ``operator'':
\begin{eqnarray}
\label{h-ansatz}
H_k = H^{(0)}_k +{\cal E}\left(ie\frac{\partial}{\partial k}\right)\;.
\end{eqnarray}
This suggests Eq.~\ref{h-ansatz} as an {\it ansatz} for the cell-periodic
Hamiltonian, now including the perturbation term $U^{pert}_k =
i\!~e{\cal E}\frac{\partial}{\partial k}$. 

In subsection~\ref{subsec:position}, we show 
that the perturbation Hamiltonian operator
given in Eq.~\ref{E-def-expand} by its matrix representation,
\begin{eqnarray}
\label{U-matrix}
{\cal U}_{mn}(k) = \left\langle u^{(0)}_{mk}\left| 
\left(i\frac{\partial}{\partial k}\right) 
\right|u^{(0)}_{nk}\right\rangle\;,
\end{eqnarray}
is the {\it periodic} part of the $\hat{x}$ operator.

Another way of arriving at this ansatz is by the following
argument. Consider the first-order change in the total energy which
can be obtained, without postulating the existence of the
field-dependent functions, by combining Eqs.~\ref{P-bphase} and
\ref{E-nv}, as follows:
\begin{eqnarray}
\label{E1-def}
E^{(1)} = - a P^{(0)} {\cal E} = \frac{a}{\pi}
\int_{0}^{\frac{2\pi}{a}}dk \sum_{n=1}^N {\cal E} 
\left\langle u_{nk}^{(0)}\left|
\left(ie\frac{\partial}{\partial k}\right)\right| u_{nk}^{(0)}
\right\rangle\;.
\end{eqnarray}
This is simply the coupling of the spontaneous polarization of the
system to the external field. From textbook perturbation theory, the
first order change in the energy is given by the diagonal matrix
elements of the perturbation term, leading again to the form of the
Hamiltonian in Eq.~\ref{h-ansatz}.

Some remarks are needed about the application of perturbation theory
to this Hamiltonian. Strictly speaking, ${\cal U}_{mn}(k)$ is not an
operator (its transformation properties under unitary transformations
will be discussed in subsection~\ref{subsec:position}). 
However, it can be
shown~\cite{nunes} that this is reflected only in the first-order
change of the single-particle eigenvalues, when we analyze
single-particle quantities obtained in the perturbation expansion of
Eq.~\ref{h-ansatz}. All other single-particle quantities, such as
wave-function derivatives and higher-order eigenvalue derivatives are
actually gauge-invariant, and thus well defined. More importantly, in
the following developments we will be interested only in quantities 
that are {\it integrated} over the Brillouin zone 
(the derivatives of the total-energy with
respect to the applied field), which will be shown to be
gauge invariant. As a final observation, it can also be
shown~\cite{nunes} that under unitary transformations, the first-order
eigenvalue acquires a change which, when integrated over the Brillouin
zone, leads to a first-order energy derivative which is defined modulo
the quantity $-e{\cal E} \ell$, where $\ell = N a$ is a lattice vector
($N$ is an integer). This is consistent with the fact that, in the
MTP, the zero-field polarization itself is defined modulo $-e\ell$.

The continuous formulation of our theory is based on the application
of a variational perturbation treatment to Eqs.~\ref{E-def} and
\ref{h-ansatz}. Alternatively, the PEAD formulation is derived by applying
the variational principle to the total energy written in terms of the
discretized form of the polarization. In this case, we combine
Eqs.~\ref{P-discr} and \ref{E-nv} to write 
\begin{eqnarray}
\label{E-def-P}
E\left[\left\{u_{nk_j}\right\};{\cal E}\right] = \frac{2}{N_k}
\left\{\sum_{n=1}^N \sum_{j=1}^{N_k} \left\langle u_{nk_j} 
\left| H^{(0)}_{k_j} \right| u_{nk_j} \right\rangle 
- \frac{e{\cal E}}{\Delta k} \sum_{j=1}^{N_k} {\rm Im} \left\{
\ln \det \left[ S_{mn}(k_j,k_{j+1}) \right] \right\}\right\}\;,
\end{eqnarray}
where $j$ runs over the $N_k$ $k$-vectors in the discretized Brillouin
zone, $\Delta k = 2\pi/a N_k$, and
\begin{eqnarray}
\label{S-matrix}
S_{nm}(k_j,k_{j+1}) = \left\langle u_{nk_j} \left|~\!
u_{mk_{j+1}}\right\rangle\right. 
\end{eqnarray}
is the overlap matrix between states at adjacent points in the
reciprocal-space mesh.  


\subsection{Position operator for periodic systems}
\label{subsec:position}

In view of the above discussion, we examine now the action of the
position operator in a space of periodic functions. Keeping in
mind that we wish to retain the periodicity of the charge density, we
seek to arrive at a consistent definition for the action of ${\hat{x}}$
in that  space. This problem has been recently tackled by
Resta~\cite{resta4}, who suggested an intrinsically many-body
redefinition of ${\hat{x}}$, in the context of periodic systems.  In
the spirit of retaining a single-particle picture, here we only offer
a heuristic justification for the form of the perturbation term given
in Eq.~\ref{h-ansatz}. For this, we use the crystal momentum
representation (CMR), following the discussion in
the paper by Blount.~\cite{blount}

Let $f(x)$ denote a square-integrable
function. The full set of zero-field Bloch eigenstates of a
periodic Hamiltonian forms a complete basis to expand $f(x)$:
\begin{equation}
\label{fx}
f(x) = \frac{1}{2\pi} \int dk \sum_n \psi_{nk}^{(0)}(x) f_n(k) = \frac{1}{2\pi}
\int dk\,e^{i kx} \sum_n u_{nk}^{(0)}(x) f_n(k) \;.
\end{equation}

The action of ${\hat{x}}$ on $f(x)$ is given in the CMR by the
expression
\begin{equation}
\label{x_op}
{\hat{x}}f(x) = \frac{1}{2\pi} \int dk\,e^{i kx} \sum_n u_{nk}^{(0)}(x) \left[
i \frac{\partial f_n(k)}{\partial k} + \sum_{n^\prime} {\cal
U}_{n n^\prime}(k) f_{n^\prime}(k) \right]\;.
\end{equation}
where ${\cal U}_{n n^\prime}$ is defined in Eq.~\ref{U-matrix}. 

Blount examined the transformation properties of the two terms
appearing in Eq.~\ref{x_op}, with respect to the choice of the phases
of the Bloch orbitals. Consider the following CMR decomposition of ${\hat{x}}$:
\begin{equation}
\label{xdecomp}
{\hat{x}} = i \frac{\partial }{\partial k} + {\cal U}_{n n^\prime}(k) =
x_d + {\cal U}_{n n^\prime}(k)\;,
\end{equation}
where $x_d = i \frac{\partial }{\partial k}$ is diagonal in the band
index. He showed that when the Bloch orbitals are multiplied by a
phase factor $e^{i\phi_n(k)}$, the term $x_d$ transforms as
$x_d^\prime = x_d -\delta_{n n^\prime} \partial\phi_n(k)/\partial k$,
while a compensatory change occurs in the diagonal term ${\cal
U}_{nn}$. So, the two terms in Eq.~\ref{xdecomp} do not transform
separately like operators, while their sum does. In our formulation,
we use the second term on the right, ${\cal U}_{nn^\prime}$, to define
a periodic Hamiltonian for the electric-field problem which, from this
discussion, is not by itself an operator in the strict sense.

We show now that ${\cal U}_{n n^\prime}(k)$ is translationally
invariant, while $x_d$ (like ${\hat x}$) is not.  Consider a
translation $T_\ell$ by a lattice vector $\ell$. The commutation
relation for ${\hat{x}}$ and $T_\ell$ is written.
\begin{equation}
\label{xcommut}
\left[{\hat{x}},T_\ell\right] = \ell T_\ell\;.
\end{equation}

To obtain the commutation relation of the perturbation term in
Eq.~\ref{h-ansatz}, we expand Eq.~(\ref{xcommut}) in the CMR
representation. Let $g(x) = T_\ell f(x) = f(x-\ell)$. From
Eq.~(\ref{fx}) and $u_{nk}^{(0)}(x-\ell) = u_{nk}^{(0)}(x)$ we get
\begin{equation}
\label{gx}
g(x) = \frac{1}{2\pi} \int dk\,e^{i k(x-\ell)} \sum_n u_{nk}^{(0)}(x)
f_n(k)\;.
\end{equation}

From this expression, it follows that
\begin{equation}
\label{gk}
g_n(k) = e^{-i k\ell} f_n(k).
\end{equation}

Further, from Eq.~(\ref{x_op}) we obtain
\begin{equation}
\label{tlxf}
T_\ell \hat{x}f(x) = \frac{1}{2\pi} \int dk\,e^{i k(x-\ell)} \sum_n
u_{nk}^{(0)}(x) \left[ i \frac{\partial f_n(k)}{\partial k} +
\sum_{n^\prime} {\cal U}_{n n^\prime}(k) f_{n^\prime}(k) \right]\;;
\end{equation}
while from Eqs.~(\ref{x_op})~and~(\ref{gk}) we get
\begin{eqnarray}
\label{xtlf}
\hat{x} T_\ell f(x) = \hat{x} g(x) &=& \frac{1}{2\pi} \int dk\,e^{i kx}
\sum_n u_{nk}^{(0)}(x) \left[ i\frac{\partial g_n(k)}{\partial k} +
\sum_{n^\prime} {\cal U}_{n n^\prime}(k) g_{n^\prime}(k) \right]\;
\nonumber\\ &=& \frac{1}{2\pi} \int dk\,e^{i k(x-\ell)} \sum_n
u_{nk}^{(0)}(x) \left[ \ell f_n(k) + i\frac{\partial f_n(k)}{\partial
k} + \sum_{n^\prime} {\cal U}_{n n^\prime}(k) f_{n^\prime}(k)
\right]\;;
\end{eqnarray}
where, in the last step, we use the result $i {\partial \over \partial
k} g_n(k) = e^{-i k\ell} \left[\ell f_n(k) + i {\partial \over
\partial k} f_n(k) \right]$. Combining
Eqs.~(\ref{tlxf})~and~(\ref{xtlf}) we arrive at the CMR expansion of
the commutation relation in Eq.~(\ref{xcommut}). Moreover, the above
development immediately shows that
\begin{eqnarray}
&\left[ x_d,T_\ell \right] = \ell T_\ell \nonumber \\ 
&\left[ {\cal U}_{n n^\prime}(k),T_\ell \right] = 0\,.
\end{eqnarray}

From the above, we observe that the perturbation term in
Eq.~\ref{h-ansatz} is invariant under lattice translations.


\subsection{Wannier-function cutoff in real space}
\label{cutoff}

The definition of an energy functional at finite fields requires
careful analysis. Because the problem does not have a ground state, a
regularization procedure is required for the definition of a
numerically-stable functional capturing the physics of the state of
the system after the electric field is turned on.

In the NV treatment of the problem in real-space, regularization is
achieved with the introduction of truncated Wannier functions, which
are constrained to zero beyond a real-space cutoff $R_c$.~\cite{nv} As
discussed by NV, in the limit $R_c \rightarrow \infty$ the functional
becomes pathological, with the property that a state having an
arbitrary value for the polarization can be constructed without
changing the value of the energy, when working at fixed polarization,
or conversely with the development of a growing (infinitely many in
the $R_c \rightarrow \infty$ limit) false local minima when working at fixed
electric field.

In order to develop this analysis on a sound mathematical basis,
one performs a Legendre transformation~\cite{Zeidler86},
from the ${\cal E}$-dependent total energy
$E[{\cal E}]$ to the $P$-dependent electric enthalpy
$\tilde E[P]$:
\begin{equation}
\tilde E[P]= \inf_{\cal E} \left\{ E[{\cal E}] + a P {\cal E} \right\}.
\end{equation}
The total energy was obtained previously [see Eq.(5)] thanks to the trial
Wannier functions,
\begin{equation}
E[{\cal E}] = \inf_{\{w\}} \left\{ E[\{w\};{\cal E}] \right\}
            = \inf_{\{w\}} \left\{ E^{(0)}[\{w\}] - a {\cal E}  P[\{w\}] \right\},
\end{equation}
while a constrained search alternatively gives its Legendre transform,
\begin{equation}
\tilde E[P] =
\inf_{\{w\} {\rm such \, that} P[\{w\}]=P} \left\{ E^{(0)}[\{w\}] \right\}.
\end{equation}
The zero-electric field total energy functional of the Wannier functions
is $E^{(0)}[\{w\}] = \sum_i \left\langle w_i \left| H^{(0)} \right| w_i
\right\rangle$.

We aim at understanding the pathologies of $E[{\cal E}]$ by examining its expression
as the inverse Legendre transform of $\tilde E[P]$,
\begin{equation}
E[{\cal E}]= \inf_P \left\{ \tilde E[P] - a P {\cal E} \right\},
\end{equation}
for which we need to characterize the minima
of $\tilde E[P]$, as well as their local behavior.

We consider, for simplicity, the case of a single occupied
band. For a given finite value of $R_c$, the electric enthalpy is a periodic
function of $P$, and $\tilde E[P_0] = E^{(0)}[\{w_0\}] = E_0$ is
the zero-field ground-state
energy ($w_0$ is the zero-field valence-band Wannier function). For
large values of $R_c$, it becomes possible to build a set of
$\ell$-dependent functions (to be normalized),
\begin{equation}
\left.\vert w \right\rangle = \left.\vert w_0 \right\rangle +
P^{1/2} \ell^{-1/2}
\left.\left| w_0^{cb}(\ell) \right\rangle\right.\;,
\end{equation}
with arbitrary value of $P$, where $w_0$ is a
zero-field valence-band Wannier function centered at the origin, and
$w_0^{cb}(\ell)$ is an empty conduction-band function centered at the
site $\ell$ within the range of $R_c$, whose coefficient is on the order
of $\ell^{-1/2}$. We consider the lattice constant $a$ to be the unit of length.
The energy for these states is
\begin{equation}
\frac
{\left\langle w \left| H^{(0)} \right| w \right\rangle }
{\left\langle w \left| w \right. \right\rangle }
=
\frac
{\left\langle w_0 \left| H^{(0)} \right| w_0 \right\rangle +
  P \ell^{-1} \left\langle w_0^{cb}(\ell)
      \left| H^{(0)} \right|  w_0^{cb}(\ell) \right\rangle   }
{1 + P \ell^{-1} }
= E_0 + P \ell^{-1} (E_{cb}-E_0)\;,
\end{equation}
where $E_{cb}$ is the expectation value of the energy for the conduction band
Wannier function.
Due to the exponential decay
of Wannier functions for insulators, the value of the
polarization is
\begin{equation}
\frac
{\left\langle w \vert x \vert w \right\rangle}
{\left\langle w \left| w \right. \right\rangle }
  \approx
\frac
{
 \left\langle w_0 \vert x \vert
 w_0 \right\rangle + P \ell^{-1} \left.\left\langle w_0^{cb}(\ell)
 \right| x \left|
 w_0^{cb}(\ell) \right\rangle\right. }
{1 + P \ell^{-1} }
= P_0 + P \;,
\end{equation}
since $\left.\left\langle w_0^{cb}(\ell) \right| x \left|
w_0^{cb}(\ell) \right\rangle\right.=\ell+P_0^{cb}$.
In the limit $\ell \rightarrow \infty$, these wavefunctions have
an arbitrary value of $P$, and an energy infinitesimally close to $E_0$.
The $\tilde E[P]$ curve becomes
flat in this limit, and only derivatives in an infinitesimal region
around the ground-state solution remain well defined. The development
of multiple minima at finite fields corresponds to the same situation,
as a growing number of minima with energies that become degenerate in
the $R_c \rightarrow \infty$ are associated to states with different
values of polarization. No global minimum as a function of
polarization can be found. In the next subsection, we analyze the
behavior of the energy functional for a model system in reciprocal
space. We will show that the same pathology manifests itself in the limit
$\Delta k \rightarrow 0$, where $\Delta k$ is the discretization
of the mesh of $k$-points in the Brillouin zone.

\subsection{Reciprocal-space analysis of a model system}
\label{model1}

For the present analysis, as well as for the application of the
perturbation expansions to be developed in the following sections, we
chose a one-dimensional (1D) two-site periodic model defined by two
parameters, the hoping integral $t$, and the on-site term which we
choose as $-\Delta/2$ and $\Delta/2$, for sites $1$ and $2$,
respectively. The Hamiltonian can be rescaled by $\Delta$ to become a
one-parameter (${t \over \Delta} \rightarrow t$) model, defined as
\begin{eqnarray}
\label{1d-ham}
H = \sum_l \left\{ \frac{1}{2} c_{2,l}^\dagger c_{2,l} -
\frac{1}{2} c_{1,l}^\dagger c_{1,l} + t~\left[c_{1,l}^\dagger c_{2,l} +
c_{2,l}^\dagger c_{1,l+1} + h.c.\right]\right\}\;,
\end{eqnarray}
where $l$ runs over unit cells. Whenever we are concerned with the 1D
model, we will consider all distances to be rescaled by the unit-cell
period (i.e. we set $a=1$ in the present section, in
Sec.~\ref{sec:1d-model}, and in Appendix~\ref{app:1dmodel}), such that
on each cell, denoted by the integer $l$, we have the basis functions
$\phi_{1}(l)$ and $\phi_2(l+1/2)$.

We apply Bloch's theorem to write the Schr\"{o}dinger equation for the
cell-periodic functions: 
\begin{eqnarray}
\label{1d-schrd}
\left. H_k^{(0)} \right| \left. u_{nk}^{(0)}\right\rangle = 
\varepsilon_{nk}^{(0)} \left.\left| u_{nk}^{(0)} \right.\right\rangle \;,
\end{eqnarray}
where $H_k^{(0)}$ is the zero-field cell-periodic Hamiltonian. In the
basis of periodic functions $\chi_1 = \sum_l \phi_{1}(l)$ and $\chi_2
= \sum_l \phi_{2}(l+1/2)$ we have
\begin{eqnarray}
\label{1d-hmtrx}
H_k^{(0)} = \left( 
\begin{array}{cc} \displaystyle -\frac{1}{2} &2t\cos\frac{k}{2} 
\\[0.08in] 2t\cos\frac{k}{2} &\frac{1}{2}%
\end{array} 
\right) \;.
\end{eqnarray}
The corresponding secular equation, $\det \left[ H_k^{(0)} -
\varepsilon_k^{(0)} {\rm 1} \right] = 0$, is easily solved for the
eigenvalues
\begin{eqnarray}
\label{1d-eigval}
\varepsilon_k^{(0)} 
= {\Huge \vphantom{a}^+_-}\left[ \frac{1}{4} + 
4t^2\cos^2{k \over 2} \right]^{1 \over 2}
= {\Huge \vphantom{a}^+_-}\frac{1}{2}\left[ 1 + 
A^2\cos^2{k \over 2} \right]^{1 \over 2},
\end{eqnarray}
where $A=4t$. Negative and positive eigenvalues correspond to valence and
conduction bands, respectively. Because the Hamiltonian is real, we
can use the following parameterization for the corresponding
eigenstates
\begin{eqnarray}
\label{1d-eigstate}
\left.\left| u_{vk}^{(0)}\right. \right\rangle &=& \left( 
\begin{array}{c} \displaystyle 
\cos\Theta_k \\[0.05in] 
\sin\Theta_k 
\end{array} 
\right) {\rm e}^{i\alpha_{vk}} \nonumber\\
\left.\left| u_{ck}^{(0)} \right. \right\rangle &=& \left( 
\begin{array}{c} \displaystyle 
~\sin\Theta_k \\[0.05in] 
\!\!-\cos\Theta_k 
\end{array} 
\right) {\rm e}^{i\alpha_{ck}}\;,
\end{eqnarray}
where $\alpha_{vk}$ and $\alpha_{ck}$ are real numbers, with no lack
of generality.  Coming back to the eigenvalue equation $H_k^{(0)}
\left.\left| u_{vk}^{(0)}\right. \right\rangle = \varepsilon_{vk}^{(0)} 
\left.\left| u_{vk}^{(0)}\right.
\right\rangle$, we obtain
\begin{eqnarray}
\label{theta}
\tan\Theta_k = \frac{\varepsilon^{(0)}_{vk} + 
\frac{1}{2}}{2t\cos{k \over 2}}\;.
\end{eqnarray}

Integrating Eq.~\ref{1d-eigval} over the Brillouin zone gives the
energy per unit cell:
\begin{eqnarray}
\label{1d-E0}
E_0 = \frac{1}{\pi} \int_0^{2\pi} dk\!~\varepsilon_{vk}^{(0)} =
-\frac{2}{\pi} \int_0^\frac{\pi}{2} dy \left[ 1+ A^2 \cos^2 y
\right]^\frac{1}{2}\;.
\end{eqnarray}

In order to discuss the pathology of the finite electric-field
functional in $k$-space, we consider a set of trial
cell-periodic functions
\begin{eqnarray}
\left.\left| u_k \right\rangle\right. = \left( 
\begin{array}{c} \displaystyle 
\cos\Theta_k~\!e^{i\alpha_k}\\[0.05in] 
\sin\Theta_k~\!e^{i\beta_k}
\end{array} \right)\;
\end{eqnarray}
for $k \in [-\pi,\pi]$, where $\Theta_k$, $\alpha_k$, and $\beta_k$
are real numbers. Imposing the condition $\left.\left| u_{k+G}\right\rangle\right. = e^{iGr}\left.\left| u_k \right\rangle\right.$,~\cite{ksv} we obtain
\begin{eqnarray} 
\cos\Theta_{k+2\pi}~\! e^{i\alpha_{k+2\pi}} &=& \cos\Theta_{k}~\!
e^{i\alpha_k} \nonumber\\
\sin\Theta_{k+2\pi}~\! e^{i\beta_{k+2\pi}} &=& -\sin\Theta_{k}~\! 
e^{i\beta_k}\,,
\end{eqnarray}
which leads to $\alpha_{2\pi} - \alpha_0 = N_\alpha \pi$ and 
$\beta_{2\pi} - \beta_0 = N_\beta \pi$. 

The expectation value of the zero-field Hamiltonian in the set of
trial wave-functions gives
\begin{eqnarray} 
\label{e0uk}
E^{(0)}[\{u_k\}] =\frac{1}{\pi}\int_0^{2\pi} dk 
~\!\left\langle u_k \left| H_k^{(0)} 
\right| u_k \right\rangle 
= \frac{1}{\pi}\int_0^{2\pi} dk \left[ -\frac{1}{2}
\cos\left(2\Theta_k\right) + 2t
\cos\left(\frac{k}{2}\right)\sin\left(2\Theta_k\right)
\cos\gamma_k\right]\;,
\end{eqnarray}
where $\gamma_k = \alpha_k - \beta_k$. 

The polarization for the trial state is
\begin{eqnarray}         
\label{puk}
P[\{u_k\}] =  \frac{ie}{\pi}\int_0^{2\pi} dk ~\!\left\langle u_k 
\left| \frac{\partial}{\partial k} \right| u_k \right\rangle =
-\frac{e}{2\pi}\left\{ \left[\vphantom{\sum_a^b}\alpha_k +
\beta_k\right]_0^{2\pi} + \int_0^{2\pi} dk~\!
\cos\left(2\Theta_k\right) \frac{\partial\gamma_k}{\partial
k}\right\}\;.
\end{eqnarray}

Minimization of $E^{(0)}[\{u_k\}]$ with respect to $\Theta_k$ and
$\gamma_k$, by setting $\partial E^{(0)}_{k}/\partial \Theta_k = 0$ and
$\partial E_k^{(0)}/\partial \gamma_k = 0$, with 
$E_k^{(0)} = \left\langle u_k \left| H_k^{(0)} \right| u_k \right\rangle$, 
leads to
\begin{eqnarray}
\label{min1}
&&\tan (2\Theta_k) = - 4t \cos\left(\frac{k}{2}\right) \cos \gamma_k \;; \\
\label{min2}
&&\left\{
\begin{array}{c} \displaystyle 
\sin\gamma_k = 0 \\[0.02in] 
{\rm or} \\[0.02in]
2t\cos\left(\frac{k}{2}\right)\sin \left( 2\Theta_k \right) = 0 \;.
\end{array} 
 \right. 
\end{eqnarray}
At $\cos(k/2) = 0$, the solution of Eqs.~\ref{min1} and \ref{min2} leads to
$\sin(2\Theta_k) = 0$ and also implies that $\gamma_k$ is {\it undefined}.
At $\cos(k/2) \ne 0$, a minimum solution is obtained by setting 
\begin{eqnarray}
\label{min3}
&&\left\{
\begin{array}{c} \displaystyle 
\sin\gamma_k = 0 \Rightarrow \gamma_k = N_\gamma\times 2\pi\;,\\[0.02in] 
{\rm and} \\[0.02in]
\tan (2\Theta_k) = - 4t \cos\left(\frac{k}{2}\right)\;.
\end{array} 
 \right. 
\end{eqnarray}

The ground-state solution is given by $\gamma_k = 0$
(i.e., $\alpha_k = \beta_k$ as in Eq.~\ref{1d-eigstate}) for all
values of $k$, with $\Theta_k$ defined by Eq.~\ref{min3}. Note that a
solution where $\gamma_k$ jumps by a multiple of $2\pi$ at $k = {\Huge
\mbox{}^+_-}\pi$ is also consistent with
Eqs.~\ref{min1}-\ref{min3}, but not with the restriction that $u_k$ be
differentiable with respect to $k$. Note also that, due to inversion
symmetry, the zero-field ground-state polarization must vanish (modulo
$-e$). This is what is obtained from Eq.~\ref{puk}, by setting
$\partial \gamma_k/\partial k = 0$, $\alpha_k = \beta_k$, and
$\alpha_{2\pi} - \alpha_0 = N_\alpha \pi$.

We consider now a trial wavefunction where $\Theta_k$ is the same as
in the ground-state solution, while $\gamma_k$ behaves as shown in
Fig.~\ref{gamma-k}, where it jumps by a value of $2\pi$ over an
interval $\Delta k$ centered at an arbitrary value of $k$. We show now
that in the $\Delta k \rightarrow 0$ limit this function can be
tailored to give an arbitrary value of the polarization, while its
energy differs from the ground-state by an infinitesimal amount, of
order $\Delta k$. 

The change in polarization for this state, with respect to the
ground-state solution, is given by
\begin{eqnarray}
\Delta P = \frac{-e}{2\pi}\int_0^{2\pi} dk~\!\cos\left(2\Theta_k\right) 
\frac{\partial \gamma_k}{\partial k}\approx 
-e \cos \left(2\Theta_{\left\langle k\right\rangle}\right).
\end{eqnarray}
$\Delta P$ in the above equation assumes values between $-e$ and
$-e/(1+16t^2)^{1/2}$. By adding another kink in the definition of
$\gamma_k$, where this function changes by $-2\pi$, we can build a solution
having any arbitrary value of $P$ in the interval $-e~\![0,1]$.

Let us consider the change in energy of the trial state. The function
$\cos\gamma_k$, as shown in Fig.~\ref{cos-gamma}, differs from one
over a small interval of the order of $\Delta k$. The change in energy
with respect to the ground state is then
\begin{eqnarray}
\Delta E = \frac{1}{\pi}\int_0^{2\pi} dk~\!
2t\cos\left(\frac{k}{2}\right)
\sin \left(2\Theta_k\right) \left(\cos\gamma_k -1\right)\approx 
-\frac{2t}{\pi}\cos\left(\frac{\left\langle k\right\rangle}{2}\right)
\sin \left(2\Theta_{\left\langle k\right\rangle}\right)\times \Delta k.
\end{eqnarray}

So, for the trial state $\Delta E \rightarrow 0 $ when $\Delta k
\rightarrow 0$. The $E^{(0)}(P)$ curve becomes flat in this
limit. This is the same pathology as the one discussed by NV in the
real-space case. Discretization of the $k$-space mesh in the Brillouin
zone is thus essential for the numerical stability of the energy
functional.

Now, we show that for the discretized version of the formulation, a
change in $P$ implies a finite change in the energy.  The discretized
polarization is written
\begin{eqnarray}
P\left[{u_{k_j}}\right] &=& \frac{e}{\pi}\sum_{k=1}^{N_k} {\rm Im}
\left[ \ln 
\left.\left\langle u_{k_j} \right| u_{k_{j+1}}\right\rangle\right]
\nonumber\\
&=& \frac{e}{\pi}\left\{\Delta N_\beta + {\rm Im} \sum_{k=1}^{N_k} 
\ln \left[ \cos\Theta_{k_j}\cos\Theta_{k_{j+1}} 
e^{i\left(\gamma_{k_{j+1}} -\gamma_{k_j}\right)} +
\sin\Theta_{k_j}\sin\Theta_{k_{j+1}}\right]\right\}\;;
\end{eqnarray}
with the energy given by
\begin{eqnarray}
E^{(0)}\left[{u_{k_j}}\right]  =  \frac{2}{N_k} \sum_{j=1}^{N_k} \left\langle
u_{k_j} \left| H_{k_j}^{(0)}\right| u_{k_{j+1}}\right\rangle
 = \frac{2}{N_k} \sum_{j=1}^{N_k}\left[ -\frac{1}{2}
\cos \left(2\Theta_{k_j}\right) +
2t\cos\left(\frac{k_j}{2}\right)\sin \left(2\Theta_{k_j}\right)
\cos\gamma_{k_j}\right]\;.
\end{eqnarray}

Again we consider the ground-state solution $\gamma_{k_j} = 0$, with
$\Theta_{k_j}$ given by Eq.~\ref{min3}. An arbitrary change in
polarization can be introduced by setting $\gamma_{k_j} \neq 0$ at a
given $k_j$, while keeping the values of $\Theta$ and $\gamma$ at all
the other $k$-points unchanged. In this case, it can be immediately
seen that a {\it finite} change $\Delta E_0 = (4t/N_k) \cos
\left(\frac{k_j}{2}\right) \sin \left(2 \Theta_{k_j}\right)
(\cos\gamma_{k_j}-1)$ is introduced in the discretized energy.


\subsection{Summary}
\label{sec2-summ}

The theoretical treatment of a periodic insulator placed in an
homogeneous electric field is plagued by severe conceptual
difficulties: (1) the potential associated with an electric field is
non-periodic and unbounded; (2) for that reason the spectrum of
electronic states changes non-analytically upon the application of a
homogeneous electric field; (3) the quantity conjugated to the
electric field, namely the polarization, cannot be computed as the
expectation value of the position (or any other) operator; (4) local
minima of the energy functional can be defined only in an
infinitesimally small region as a function of the polarization, the
energy functional being perfectly flat otherwise.

In order to address problems 1 and 2, following Nenciu, we restrict
ourselves to periodic-polarized-insulating states, of which the lowest
in energy is a long-lived resonance of the unrestricted system.
Keeping this restriction in mind, we show that the position operator
can be decomposed, in the crystal momentum representation, into a
non-periodic part and a periodic part. The latter can be 
introduced in an ansatz Hamiltonian acting on the periodic part of
the Bloch functions, from which the Berry phase formulation of the
polarization is recovered, solving problem 3 as well.

We are aware that this line of thought does not yet justify rigorously
the use of this Hamiltonian: a more careful derivation, in the spirit
of the mathematical work of Nenciu, would be needed. However, this
rather simple Hamiltonian allows to recover all the previously known
lowest-order expressions for the polarization and its derivatives, and
to derive other low-order expressions as well as generic expressions
to all orders, as we shall see in the coming sections.

Problem 4 is solved by introducing a regularization procedure in
reciprocal space, similar in spirit to the real-space cutoff radius
introduced by NV. For the regularized energy functional, the local
minima have a finite basin of attraction as a function of the
polarization.


\section{Perturbation theory applied to a variational total-energy functional}
\label{sec:VPT}

In view of the application of perturbation theory to Eq.~\ref{E-def},
we summarize now the variational formulation of DFPT, as presented in Ref.
~\onlinecite{gonze6}. We
consider the formalism at its non-self-consistent level, without including
the Hartree and exchange-correlation terms of the perturbative
expressions.

One considers a perturbative expansion of a variational
principle applied to the electronic total-energy functional.
In terms of the small
parameter $\lambda$ associated with the perturbation, the
perturbation series reads
\begin{eqnarray}
\label{series}
{\cal O} (\lambda) &=& {\cal O}^{(0)} + \lambda {\cal O}^{(1)} +
\lambda^2 {\cal O}^{(2)} + \lambda^3 {\cal O}^{(3)} + ...\;,
\nonumber\\ {\cal O}^{(n)} &=& \left. \frac{1}{n!}\frac{d^n {\cal
O}(\lambda)}{d\lambda^n}~\!
\right|_{~\!\lambda = 0}
\end{eqnarray}
for a generic observable ${\cal O}$.
The system Hamiltonian is $H = K + v_{ext}$, and the total-energy functional is
\begin{eqnarray}
E = \sum_{\alpha=1}^N \left\langle
 \varphi_{\alpha} \left| \left( K + v_{ext}\right) \right| \varphi_{\alpha} 
\right\rangle\;,
\end{eqnarray}
where $K$ and $v_{ext}$ are the kinetic-energy and external-potential
operators. The total-energy functional is to be minimized under the
orthonormality constraints for the occupied wavefunctions
\begin{eqnarray}
\label{ortho}
\left\langle \varphi_{\alpha} \left| \varphi_{\beta} \right\rangle \right. = 
\delta_{\alpha\beta}.
\end{eqnarray}

Using the Lagrange-multiplier method, the functional
\begin{eqnarray}
\label{F-ener}
F = \sum_{\alpha=1}^N \left\langle \varphi_{\alpha}
\left| \left( T + v_{ext}\right) \right| \varphi_{\alpha}\right\rangle
-\sum_{\alpha,\beta = 1}^N \Lambda_{\beta\alpha}
\left[ \left\langle~\!\varphi_\alpha~\!\left|
~\!\varphi_\beta~\!\right\rangle \right.
- \delta_{\alpha\beta} \right]
\end{eqnarray}
is minimized with respect to the wavefunctions. The minimum condition,
$\delta F/\delta\varphi_\alpha^\ast = 0$, leads to the Euler-Lagrange
equation
\begin{eqnarray}
\label{EL}
\left. H\left.\right| \varphi_\alpha~\!\right\rangle = 
\sum_{\beta = 1}^N \left. \Lambda_{\beta\alpha}\left.
\right| ~\!\varphi_\beta~\!\right\rangle\;.
\end{eqnarray}

Eq.~\ref{EL} represents a set of generalized eigenvalue equations
which assume the form of the usual eigenvalue equations when the
so-called diagonal gauge is chosen to fix the phase arbitrariness of
the wavefunctions.~\cite{gonze6} Here, we keep the generalized form, as
needed for the choice of gauge to be used in our
theory. An expression for the Lagrange-multiplier
matrix is obtained by multiplying Eq.~\ref{EL} by an occupied
wavefunction, leading to

\begin{eqnarray}
\label{LM}
\Lambda_{\beta\alpha} =
\left\langle \varphi_\beta \left| H \right| 
\varphi_\alpha \right\rangle\;.
\end{eqnarray}

We consider now the perturbation expansion of
Eqs.~\ref{ortho}-\ref{LM}.
The orthonormalization condition becomes
\begin{eqnarray}
\label{ortho-pert}
\sum_{j = 0}^i \left. \left\langle \varphi_{\alpha}^{(j)} \right|
\varphi_{\beta}^{(i-j)} \right\rangle = 0 ~\;\;\;~{\rm for} ~\;\;\;~i\geq
1\;.
\end{eqnarray}

The expansion of Eq.~\ref{EL} gives the generalized Sternheimer equation
\begin{eqnarray}
\label{EL-pert}
\sum_{j =0}^i H^{(j)}~\!\left.\left|\varphi_\alpha^{(i-j)}~\!\right\rangle\right.
= \sum_{j=0}^i\sum_{\beta = 1}^N \Lambda_{\beta\alpha}^{(j)}
\left.\left|~\!\varphi_\beta^{(i-j)}~\!\right\rangle\right.\;,
\end{eqnarray}
where $H^{(i)} = T^{(i)} + v_{ext}^{(i)}$ is the $ith$-order term in
the expansion of the Hamiltonian.

The expansion of the Lagrange-multiplier matrix is given by
\begin{eqnarray}
\label{LM-pert}
\Lambda_{\beta\alpha}^{(i)} = \sum_{j = 0}^i \sum_{k =0}^i
\left\langle~\!\varphi_\beta^{(j)}~\!\left|~\! H^{(i-j-k)}~\!
\right|~\!\varphi_\alpha^{(k)}~\!\right\rangle \;.
\end{eqnarray}

Finally, a generic term in the perturbative expansion of the
total-energy functional in Eq.~\ref{F-ener} is written
\begin{eqnarray}
\label{E-pert}
E^{(i)} &=& \sum_{\alpha = 1}^N \sum_{l = 0}^j\sum_{k =0}^{i}
\sum_{l^\prime = 0}^j \delta (i - l - k - l^\prime)
\left\langle\varphi_{\alpha}^{(l)}\left|H^{(k)}\right|
\varphi_\alpha^{(l^\prime)}\right\rangle \nonumber \\ 
&-& \sum_{\alpha , \beta = 1}^N \sum_{l = 0}^j\sum_{k=0}^{i-j-1}
\sum_{l^\prime = 0}^j \delta (i - l - k - l^\prime)
\Lambda_{\beta\alpha}^{(k)} \left\langle\varphi_{\alpha}^{(l)}\left|
\varphi_\beta^{(l^\prime)}\right\rangle\right.\;,
\end{eqnarray}
where $i = 2 j$ or $i = 2 j + 1$. We remark that only
wavefunctions derivatives up to order $\lambda^j$ appear in the
$i$th-order term of the energy, as a result of the $2n+1$-theorem.
Moreover, a minimum principle holds for $E^{(2j)}$ with respect to the
$j$th-order variations of the wave functions, i.e., $\delta E^{(2
j)}/\delta \varphi_\alpha^{(j)} = 0$.

A particularly useful result derived in Ref.~\onlinecite{gonze6} is a set of
non-variational expressions for the second-order derivative of the
energy. In the present work, the Hamiltonian is of first-order in
the perturbation ($v^{(i)}_{ext}=0\;{\rm for}\; i\ge 2$), 
in which case the non-variational expressions are given by
\begin{eqnarray}
\label{E2-nv}
E^{(2)} &=& \sum_{\alpha = 1}^N
\left\langle\varphi_{\alpha}^{(1)}\left|v_{ext}^{(1)}\right|
\varphi_\alpha^{(0)}\right\rangle = \sum_{\alpha = 1}^N
\left\langle\varphi_{\alpha}^{(0)}\left|v_{ext}^{(1)}\right|
\varphi_\alpha^{(1)}\right\rangle \nonumber \\
 &=& \sum_{\alpha = 1}^N
\frac{1}{2}\left\langle\varphi_{\alpha}^{(1)}\left|v_{ext}^{(1)}\right|
\varphi_\alpha^{(0)}\right\rangle +
\frac{1}{2}\left\langle\varphi_{\alpha}^{(0)}\left|v_{ext}^{(1)}\right|
\varphi_\alpha^{(1)}\right\rangle\,.
\end{eqnarray}

The zeroth-order wave functions are chosen to obey the unperturbed eigenvalue
equation $H^{(0)}\left.\left| \varphi^{(0)}_{\alpha} \right.\right\rangle =
\varepsilon^{(0)}_{\alpha}\left.\left| \varphi^{(0)}_{\alpha} \right.\right\rangle $. From
Eq.~\ref{LM-pert}, the zeroth-order Lagrange-multiplier matrix is
given by
\begin{eqnarray}
\label{lambda0}
\Lambda_{\beta\alpha}^{(0)} = \delta_{\beta\alpha}\varepsilon_\alpha^{(0)}\;.
\end{eqnarray}

In the present work, we use the so-called ``parallel-transport''
gauge, as discussed in Ref.~\onlinecite{gonze6}. In this gauge, 
the following condition is imposed on the derivatives of the
wave functions
\begin{eqnarray}
\label{diag-gauge}
\left.\left\langle~\!\varphi_\alpha^{(0)}~\!\right|~\!\varphi_\beta^{(i)}~\!\right\rangle
-\left.\left\langle~\!\varphi_\alpha^{(i)}~\!\right|~\!\varphi_\beta^{(0)}~\!\right\rangle
= 0\;;
\end{eqnarray}
which allows us to rewrite the expansion of the orthonormalization
condition as
\begin{eqnarray}
\label{ortho-pert1}
\left.\left\langle~\!\varphi_\alpha^{(0)}~\!\right|~\!\varphi_\beta^{(i)}
~\!\right\rangle = \left\{
\begin{array}{lcc} \displaystyle -\frac{1}{2}
\sum_{j=1}^{i-1} \left. \left\langle~\!\varphi_{\alpha}^{(j)}~\!\right|
~\!\varphi_{\beta}^{(i-j)}\right\rangle &{\rm for} &i > 1\;;\\[0.2in] ~\;~0
&{\rm for} &i = 1\;. \end{array}
\right.
\end{eqnarray}


\section{Perturbation theory applied to the continuous form}
\label{sec:pt-cont}


\subsection{Perturbation expansion and proof of gauge invariance}
\label{expan-cont}

Following the discussion in Sec.~\ref{sec:efp}, we can develop a
perturbation expansion for the electric-field problem.  In this
section, we discuss the continuous form of the theory. The
cell-periodic Hamiltonian, including the perturbation term, is given
in Eq.~\ref{h-ansatz}. We apply the machinery of the variational DFPT
to this Hamiltonian, by postulating that the
expression
\begin{eqnarray}
\label{E0}
E\left[\left\{u_{nk}\right\};{\cal E}\right] = {a \over
\pi} \int_0^{\frac{2\pi}{a}}dk \left[ \sum_{n=1}^N \left\langle u_{nk} 
\left| H_k^{(0)} + i e {\cal E} \frac{\partial}{\partial k} \right|
u_{nk}\right\rangle \right]\;
\end{eqnarray}
is to be minimized with respect to the $\left\{u_{nk}\right\}$, under
the constraints
\begin{equation}
\label{ortho0}
\left.\left\langle u_{mk} \right| u_{nk}\right\rangle = \delta_{mn}\;.
\end{equation}
A local minimum will exist for the functional in Eq.~\ref{E0} provided
that a discretization of the $k$-space integrals is performed. The
continuum formulation which is considered in this section is valid only at
infinitesimal fields.

Applying Eq.~\ref{F-ener}, we write
\begin{equation}
\label{E1}
F\left[ \left\{u_{nk}\right\};{\cal E}\right] = {a \over
\pi} \int_0^{\frac{2\pi}{a}}dk \left[ \sum_{n=1}^N \left\langle u_{nk} 
\left|
H_k^{(0)} + i e {\cal E} \frac{\partial}{\partial k} \right|
u_{nk}\right\rangle - \sum_{m,n=1}^N \left\{
\left.\left\langle u_{nk} \right| u_{mk}\right\rangle 
- \delta_{nm}\right\} \Lambda_{mn}(k) \right]\;.
\end{equation}
The unconstrained minimization of this functional is obtained by
setting $\delta F\left[ \left\{u_{nk}\right\}\right]/\delta u_{nk}$ =
0, leading to the corresponding Euler-Lagrange equation
\begin{eqnarray}
\label{stern0}
\left( H_k^{(0)} +i e {\cal E} \frac{\partial}{\partial k}
\right) \left.\left| u_{nk}\right\rangle \right. 
- \sum_{m=1}^N \Lambda_{mn}(k) \left.
\left| u_{mk}\right\rangle\right. = 0\;.
\end{eqnarray}

Next, we consider separately the perturbation expansions of
Eqs.~\ref{E1} and \ref{stern0}. In both cases, we will demonstrate
explicitly that the general expansion term transforms properly under a
general unitary transformation of the occupied orbitals.


\subsubsection{Lagrange multipliers and orthonormalization constraints}

In the present case, Eq.~\ref{ortho-pert1} for the orthonormalization
constraints reads
\begin{eqnarray}
\label{u-ortho}
\left.\left\langle u_{m k}^{(0)} \right| u_{n k}^{(i)} \right\rangle 
= \left\{ \begin{array}{lc}
\displaystyle
-\frac{1}{2} \sum_{j=1}^{i-1} \left.\left\langle u_{mk}^{(j)} 
\right| u_{nk}^{(i-j)} 
\right\rangle\;,\;\;\;\;&i > 1\;;\\[0.2in]
\;\;\;0\;,        &i = 1\;; \end{array} 
\right.
\end{eqnarray}
giving the occupied-subspace projection of $u_{n k}^{(i)}$ in
terms of the lower-order solutions for the periodic functions.
 
Since $H^{(i)}_k\equiv 0$ for all $i \ge 2$, and $H^{(1)}_k =
ie\frac{\partial}{\partial k}$, the expansion of the
Lagrange multipliers becomes
\begin{eqnarray}
\label{LM_exp}
\Lambda_{mn}^{(i)}(k) =\sum_{j=0}^i \left\langle u_{mk}^{(j)} \left|
H_k^{(0)} \right| u_{nk}^{(i-j)}\right\rangle + \sum_{j=0}^{i-1}
\left\langle u_{mk}^{(j)} \left| ie\frac{\partial}{\partial k} \right|
u_{nk}^{(i-j-1)}\right\rangle\;.
\end{eqnarray}

In the following development, we will make explicit use of the
expressions for $\Lambda_{mn}^{(0)}(k)$, $\Lambda_{mn}^{(1)}(k)$, and
$\Lambda_{mn}^{(2)}(k)$. From Eq.~\ref{lambda0},
$\Lambda_{mn}^{(0)}(k) = \varepsilon^{(0)}_{nk}\delta_{mn}$.  
Since $\left.\left\langle
u_{mk}^{(0)} \right| u_{nk}^{(1)}\right\rangle = 0$ from Eq.~\ref{u-ortho}, 
and $\left. H_k^{(0)} \right| \left. u_{nk}^{(0)}\right\rangle = 
\left.\varepsilon_{nk}^{(0)} \right| 
\left. u_{nk}^{(0)}\right\rangle$, $\Lambda_{mn}^{(1)}(k)$ is given by
\begin{eqnarray}
\label{LM1}
\Lambda_{mn}^{(1)}(k) = \left\langle u_{mk}^{(0)} \left|
ie\frac{\partial}{\partial k} \right| u_{nk}^{(0)}\right\rangle \;.
\end{eqnarray}
The second-order term reads
\begin{eqnarray}
\label{LM2}
\Lambda_{mn}^{(2)}(k) &=&
\left\langle u_{mk}^{(0)} \left| H_k^{(0)} \right| u_{nk}^{(2)}\right\rangle + 
\left\langle u_{mk}^{(2)} \left| H_k^{(0)} \right| u_{nk}^{(0)}\right\rangle +
\left\langle u_{mk}^{(1)} \left| H_k^{(0)} \right| u_{nk}^{(1)}\right\rangle \nonumber\\
&& +~\!\left\langle u_{mk}^{(0)} \left| ie\frac{\partial}{\partial k} 
\right| u_{nk}^{(1)}\right\rangle +
\left\langle u_{mk}^{(1)} \left| ie\frac{\partial}{\partial k} 
\right| u_{nk}^{(0)}\right\rangle \;.
\end{eqnarray}
%


\subsubsection{Energy}
\label{pt-cont-energy}

The perturbation expansion for the energy is obtained from
Eq.~\ref{E-pert}. We analyze even and odd
terms separately. For the even-order terms we write
\begin{eqnarray}
\label{E-even}  
E^{(2i)} &=& \frac{a}{\pi}\int_0^{\frac{2\pi}{a}} dk
\left[\sum_{n=1}^N \left\langle u_{nk}^{(i)} \left| H_k^{(0)} \right|
u_{nk}^{(i)}\right\rangle + \left\langle u_{nk}^{(i-1)} \left|
ie\frac{\partial}{\partial k} \right| u_{nk}^{(i)}\right\rangle +
\left\langle u_{nk}^{(i)} \left| ie\frac{\partial}{\partial k} \right|
u_{nk}^{(i-1)}\right\rangle\right.  \nonumber \\ &&\left.- \sum_{m,n=1}^N
\sum_{j,j^\prime=1}^i \sum_{l=0}^{i-1} \delta(2i-j-j^\prime-l)
\Lambda_{mn}^{(l)}(k) \left\langle u_{nk}^{(j)} \left|
u_{mk}^{(j^\prime)}\right\rangle\right.\right] \;,
\end{eqnarray}
while the odd terms are given by
\begin{equation}
\label{E-odd}  
E^{(2i+1)} = \frac{a}{\pi} \int_0^{\frac{2\pi}{a}}dk
\left[\sum_{n=1}^N \left\langle u_{nk}^{(i)} \left|
ie\frac{\partial}{\partial k} \right| u_{nk}^{(i)}\right\rangle -
\sum_{n,m=1}^N \sum_{j,j^\prime,l=1}^i \delta(2i+1-j-j^\prime-l)
\Lambda_{mn}^{(l)}(k) \left\langle u_{nk}^{(j)} \left|
u_{mk}^{(j^\prime)}\right\rangle\right.\right] \;.
\end{equation}

An important aspect concerns the invariance of these expressions with
respect to the choice of phases of the Bloch orbitals.  More
generally, we must consider unitary transformations that keep the
subspace of occupied states invariant.  We show in Appendix~\ref{app:gauge} that
Eqs.~\ref{E-even} and \ref{E-odd} can be rewritten in such a way as to
display the required gauge-invariance property explicitly. The
lower-order derivatives are usually of more practical interest, and
for that reason the invariant form of the energy terms up to fourth
order are written explicitly here, along with the general
expansion term.

The second-order energy derivative is obtained by setting $i=1$ in
Eq.~\ref{E-even}. After some manipulation, this quantity can be
written in the following form:
\begin{eqnarray}
\label{E2-invar}
E^{(2)} &=& \frac{a}{\pi} \int_0^{\frac{2\pi}{a}}dk \sum_{n=1}^N
\left[ 
\left\langle u_{nk}^{(1)} \left| 
\left( H_k^{(0)} - \varepsilon^{(0)}_{nk}\right) 
\right| u_{nk}^{(1)}\right\rangle + 
\left.\left\langle u_{nk}^{(1)}\right. \right| 
\left(
ie\frac{\partial}{\partial k} \sum_{m=1}^N 
\left| \left. u_{mk}^{(0)}\right\rangle\right. 
\left.\left\langle u_{mk}^{(0)} \right|\right. 
\right) 
\left| \left. u_{nk}^{(0)}\right\rangle \right.
\right. 
\nonumber \\ 
&~& - \left.
\left.\left\langle u_{nk}^{(0)}\right. \right| 
\left(
ie\frac{\partial}{\partial k} \sum_{m=1}^N 
\left| \left. u_{mk}^{(0)}\right\rangle\right. 
\left.\left\langle u_{mk}^{(0)} \right|\right. 
\right) 
\left| \left. u_{nk}^{(1)}\right\rangle \right.
\right]\,. 
\end{eqnarray}
As expected, our formula for $E^{(2)}$ is identical to the
linear-response expression.~\cite{gonze5} 

The non-variational expression for $E^{(2)}$ (see Eq.~\ref{E2-nv}) is given by
\begin{eqnarray}
\label{E2-invar-nv}
E^{(2)} &=& -\frac{a}{\pi} \int_0^{\frac{2\pi}{a}}dk 
\sum_{n=1}^N
\left\langle u_{nk}^{(1)} \left| 
\left( H_k^{(0)} - \varepsilon^{(0)}_{nk} \right) 
\right| u_{nk}^{(1)}\right\rangle \nonumber\\
&=& \frac{a}{2\pi} \int_0^{\frac{2\pi}{a}}dk 
\sum_{n=1}^N
\left.\left\langle u_{nk}^{(1)}\right. \right| 
\left(
ie\frac{\partial}{\partial k} \sum_{m=1}^N 
\left| \left. u_{mk}^{(0)}\right\rangle\right. 
\left.\left\langle u_{mk}^{(0)} \right|\right. 
\right) 
\left| \left. u_{nk}^{(0)}\right\rangle \right. \nonumber\\
&~&- \left.\left\langle u_{nk}^{(0)}\right. \right| 
\left(
ie\frac{\partial}{\partial k} \sum_{m=1}^N 
\left| \left. u_{mk}^{(0)}\right\rangle\right. 
\left.\left\langle u_{mk}^{(0)} \right|\right. 
\right) 
\left| \left. u_{nk}^{(1)}\right\rangle \right.\,.
\end{eqnarray}

The fourth-order energy term is written
\begin{eqnarray}
\label{E4-invar}
E^{(4)} &=& \frac{a}{\pi} \int_0^{\frac{2\pi}{a}}dk \sum_{n=1}^N
\left[ \left\langle u_{nk}^{(2)} 
\left| \left( H_k^{(0)} - \varepsilon^{(0)}_{nk}
\right) \right| u_{nk}^{(2)}\right\rangle + 
\left.\left\langle u_{nk}^{(2)}\right. \right| \left(
ie\frac{\partial}{\partial k} \sum_{m=1}^N \left|
\left. u_{mk}^{(1)}\right\rangle\right. \left.\left\langle 
u_{mk}^{(0)} \right|\right. 
\right) \left|
u_{nk}^{(0)}\right\rangle \right. \nonumber \\ 
&~& - \left.
\left.\left\langle u_{nk}^{(0)}\right. \right| \left(
ie\frac{\partial}{\partial k} \sum_{m=1}^N \left|
\left. u_{mk}^{(0)}\right\rangle\right. \left.\left\langle 
u_{mk}^{(1)} \right|\right. 
\right) \left. \left|
u_{nk}^{(2)}\right\rangle \right.
\right] \,.
\end{eqnarray}
It is worth pointing out the simplicity of the expression for
$E^{(4)}$, which mirrors that of $E^{(2)}$ almost exactly.

The general even-order energy term for $i>2$ is written
\begin{eqnarray}
\label{E-even-invar}
E^{(2i)} &=& \frac{a}{\pi} \int_0^{\frac{2\pi}{a}}dk \sum_{n=1}^N
\left\{ 
\left\langle u^{(i)}_{nk} \left| 
\left( H_k^{(0)} - \varepsilon^{(0)}_{nk} \right) 
\right| u^{(i)}_{nk} \right\rangle 
+ \left. \left\langle u^{(i)}_{nk} \right. \right| 
\left(
ie\frac{\partial}{\partial k} \sum_{m=1}^N 
\left. \left| u^{(i-1)}_{mk} \right. \right\rangle 
\left. \left\langle u^{(0)}_{mk} \right. \right| 
\right)
\left. \left| u^{(0)}_{nk} \right. \right\rangle
\right. \nonumber \\ 
&~& - \left. \left\langle u^{(0)}_{nk} \right. \right|
\left(
ie\frac{\partial}{\partial k} \sum_{m=1}^N 
\left. \left| u^{(0)}_{mk} \right. \right\rangle 
\left. \left\langle u^{(i-1)}_{mk} \right. \right|
\right) 
\left. \left| u^{(i)}_{nk} \right. \right\rangle 
+ \sum_{j,j^{\prime}=1}^i \sum_{l=2}^{i-1} \delta(2i-j-j^\prime-l) 
\times \nonumber\\
&~&
\left. \left[
\sum_{l^\prime=0}^{l} \sum_{m=1}^N 
\left\langle u^{(l^\prime)}_{mk} \left| H_k^{(0)} \right|  
u^{(l-l^\prime)}_{nk} \right\rangle
\left\langle u^{(j)}_{nk} \left| u^{(j^\prime)}_{mk} \right. \right\rangle 
- \sum_{l^\prime=0}^{l-1} \left\langle \left. u^{(j)}_{nk} \right| \right. 
\left(
ie\frac{\partial}{\partial k} \sum_{m=1}^N 
\left| \left. u^{(j^\prime)}_{mk} \right\rangle \right. 
\left\langle u^{(l^\prime)}_{mk}\right. 
\right) 
\left| \left. u^{(l-l^\prime-1)}_{nk} \right\rangle \right. 
\right] \right\}\,.\nonumber\\
\end{eqnarray}

In the above and the following expressions, we use the notation
$\left(ie\frac{\partial}{\partial k} \left| \left. u_{mk} \right.
\right\rangle \left. \left\langle u_{mk} \right. \right| \right)$ to
indicate that $\partial/\partial k$ acts only on the
quantities embraced in parenthesis.  In order to demonstrate that the
energy derivatives fulfill the gauge-invariance requirement, we
consider a general gauge transformation~\cite{resta1} among the
occupied states at each k-point:
\begin{eqnarray}
\label{gauge}
\left.\left| {\tilde u}_{n k} \right.\right\rangle = 
\sum_{m} U_{mn}(k) \left.\left| u_{n k} \right. \right\rangle\;,
\end{eqnarray}
where $U$ is an unitary transformation, i.e.,
$U U^{\dagger} = 1$.
It follows immediately that
\begin{eqnarray}
\label{gauge-invar} 
\frac{\partial}{\partial k} 
\left( 
\sum_n 
\left.\left| {\tilde u}_{nk}^{(i)} \right. \right\rangle 
\left\langle \left. {\tilde u}_{nk}^{(j)} \right| \right.
\right) 
=\frac{\partial}{\partial k} 
\left( 
\sum_{lm} \sum_n U_{nl}(k) U^{\ast}_{nm}(k) 
\left.\left| u_{lk}^{(i)} \right. \right\rangle 
\left\langle \left. u_{mk}^{(j)} \right| \right.
\right) 
= \frac{\partial}{\partial k} 
\left( 
\sum_{n} \left.\left| u_{nk}^{(i)} \right. \right\rangle 
\left\langle \left. u_{nk}^{(j)} \right| \right.
\right)\;.
\end{eqnarray}

Since $\frac{\partial}{\partial k}$ acts only on gauge-invariant
quantities, Eqs.~\ref{E2-invar}-\ref{E-even-invar} are themselves
gauge-invariant. The same argument holds for the odd-order
derivatives we derive below.

Turning now to the odd-order derivatives of the energy, we set $i=1$
in Eq.~\ref{E-odd} to write the third-order term as
\begin{eqnarray}
\label{E3-invar}  
E^{(3)} = \frac{a}{\pi} \int_0^{\frac{2\pi}{a}}dk \sum_{n=1}^N
\left.\left\langle u_{nk}^{(1)}\right. \right| 
\left(
ie\frac{\partial}{\partial k} \sum_{m=1}^N 
\left| \left. u_{mk}^{(1)} \right\rangle \right. 
\left.\left\langle u_{mk}^{(0)} \right.\right| 
\right) 
\left| \left. u_{nk}^{(0)}\right\rangle \right. \,.
\end{eqnarray}
This expression for $E^{(3)}$ is identical to the one previously
derived by Dal Corso and Mauri.~\cite{dalcorso1}

The general odd-order term for $i>1$ is written
\begin{eqnarray}
\label{E-odd-invar}  
E^{(2i+1)} &=& \frac{a}{\pi} \int_0^{\frac{2\pi}{a}}dk \sum_{n=1}^N
\left\{ 
\left. \left\langle u^{(i)}_{nk} \right. \right| 
\left(
ie\frac{\partial}{\partial k} \sum_{m=1}^N 
\left. \left| u^{(i)}_{mk} \right. \right\rangle 
\left. \left\langle u^{(0)}_{mk} \right. \right| 
\right)
\left. \left| u^{(0)}_{nk} \right. \right\rangle
\right. \nonumber \\ 
&~& + \sum_{j,j^{\prime}=1}^i \sum_{l=2}^{i} 
\delta(2i+1-j-j^\prime-l) 
\left[
\sum_{l^\prime=0}^{l} \sum_{m=1}^N 
\left\langle u^{(l^\prime)}_{mk} 
\left| H_k^{(0)} \right|  
u^{(l-l^\prime)}_{nk} \right\rangle
\left\langle u^{(j)}_{nk} \left| u^{(j^\prime)}_{mk} \right. \right\rangle 
\right.
\nonumber\\
&~&-\left.\left. \sum_{l^\prime=0}^{l-1} 
\left\langle \left. u^{(j)}_{nk} \right| \right. 
\left(
ie\frac{\partial}{\partial k} \sum_{m=1}^N 
\left| \left. u^{(j^\prime)}_{mk} \right\rangle \right. 
\left\langle u^{(l^\prime)}_{mk}\right. 
\right) 
\left| \left. u^{(l-l^\prime-1)}_{nk} \right\rangle \right. 
\right] \right\}\,.
\end{eqnarray}
In appendix~\ref{app:gauge}, we demonstrate how
Eqs.~\ref{E2-invar}-\ref{E-even-invar}, \ref{E3-invar}, and
\ref{E-odd-invar} are obtained from Eqs.~\ref{E-even} and
\ref{E-odd}.


\subsubsection{Sternheimer equation}
\label{pt-cont-stern}

The projection of the wave functions on the subspace of occupied
unperturbed states is given by Eq.~\ref{u-ortho}. The projection onto
the subspace of unoccupied states is given by the projection of the
Sternheimer equation in that subspace.
The perturbation series for the Sternheimer equation can be obtained
either by expanding Eq.~\ref{stern0}, or more directly from
Eq.~\ref{E-even} above, by setting $\delta E^{(2i)}/\delta u_{n
k}^{\ast(i)} = 0$. The general expansion term is written
\begin{eqnarray}
\label{stern.i}
P_{c k} \left( H_k^{(0)} - \varepsilon_{n k}^{(0)}\right) 
P_{c k} \left.\left|
u_{n k}^{(i)} \right. \right\rangle = 
- i e P_{c k} \frac{\partial}{\partial k} 
\left.\left| u_{n k}^{(i-1)}\right. \right\rangle 
+ \sum_{m=1}^{N} \sum_{j=1}^{i-1}
\Lambda_{mn}^{(j)}(k) 
P_{c k} \left.\left| u_{m k}^{(i-j)}\right. \right\rangle\;.
\end{eqnarray}
where $P_{ck}$ is the projector onto the subspace of unoccupied
unperturbed states. This equation can be solved for
$u_{n k}^{(i)}$, once the lower-order derivatives of $u_{n k}$ and
$\Lambda_{mn}(k)$ have been obtained.

Using the invariant form for the even terms of the energy,
Eqs.~\ref{E2-invar}-\ref{E-even-invar}, we set $\delta E^{(2i)}/\delta
u_{n k}^{\ast(i)} = 0$ to obtain the explicitly invariant form of
the Sternheimer equation. For the $i=1$ and $i=2$ terms we obtain
\begin{eqnarray}
\label{stern1-invar}
P_{c k} \left( H_k^{(0)} - \varepsilon_{n k}^{(0)}\right) 
P_{c k} \left.\left| u_{n k}^{(1)}\right. \right\rangle = - P_{c k} 
\left( 
i e \frac{\partial}{\partial k} \sum_{m=1}^N 
\left.\left| u_{m k}^{(0)}\right. \right\rangle
\left\langle \left. u_{m k}^{(0)} \right|\right. 
\right) 
\left|\left. u_{n k}^{(0)} \right\rangle\right. \,,\\
\nonumber\\
\label{stern2-invar}
P_{c k} \left( H_k^{(0)} - \varepsilon_{n k}^{(0)}\right) 
P_{c k} \left.\left| u_{n k}^{(2)}\right. \right\rangle = - P_{c k} 
\left( 
i e \frac{\partial}{\partial k} \sum_{m=1}^N 
\left.\left| u_{m k}^{(1)}\right. \right\rangle
\left\langle \left. u_{m k}^{(0)} \right|\right. 
\right) 
\left|\left. u_{n k}^{(0)} \right\rangle\right. \,.
\end{eqnarray}

For the complete specification of $u_{nk}^{(1)}$ and $u_{nk}^{(2)}$, 
the projections onto the unperturbed occupied subspace are obtained
from Eq.~\ref{u-ortho}:
\begin{eqnarray}
\label{Pv-u1}
P_{vk}\left.\left| u_{nk}^{(1)}\right.\right\rangle &=& 0 \;,\\
\label{Pv-u2}
P_{vk}\left.\left| u_{nk}^{(2)}\right.\right\rangle &=& 
-\frac{1}{2}\sum_{m=1}^N 
\left.\left| u_{mk}^{(0)}\right.\right\rangle 
\left\langle\left. u_{mk}^{(1)} \right| u_{nk}^{(1)}\right\rangle\;;
\end{eqnarray}
where $P_{vk}$ is the projection operator for the occupied states.

The higher-order terms for the Sternheimer equation are given by
\begin{eqnarray}
\label{stern-invar}
P_{c k} 
\left( H_k^{(0)} - \varepsilon_{n k}^{(0)} \right) 
P_{c k} 
\left.\left| u_{n k}^{(i)}\right.\right\rangle 
= &-& P_{c k} 
\left[ 
\sum_{j=1}^{i-1} \sum_{l=0}^{j-1} 
\left(
i e \frac{\partial}{\partial k} \sum_{m=1}^N 
\left.\left| u_{m k}^{(i-j)}\right. \right\rangle 
\left\langle \left. u_{m k}^{(l)} \right|\right. 
\right) 
\left|\left. u_{n k}^{(j-l-1)} \right\rangle \right.
\right. 
\nonumber\\
&-& \left.\sum_{j=2}^{i-1} \sum_{l=0}^j \sum_{m=1}^N 
\left.\left| u_{mk}^{(i-j)}\right.\right\rangle 
\left\langle u_{m k}^{(l)} \left| H_k^{(0)} \right| 
u_{nk}^{(j-l)} \right\rangle 
\right]\;\;\;,\;\;\;i>2.
\end{eqnarray}

The valence-band component of $\left.\left| u_{n
k}^{(i)}\right.\right\rangle$ is given by
\begin{eqnarray}
\label{Pv-uvi}
P_{vk} \left.\left| u_{n k}^{(i)}\right.\right\rangle = 
-\frac{1}{2}\sum_{j=1}^{i-1} 
\left.\left| u_{n k}^{(0)}\right.\right\rangle 
\left\langle u_{mk}^{(j)} \left| u_{nk}^{(i-j)}\right.\right\rangle\;.
\end{eqnarray}
As in the case of the energy terms, in Eqs.~\ref{stern1-invar},
\ref{stern2-invar}, and \ref{stern-invar} the derivative $\partial/\partial
k$ acts only on gauge-invariant quantities.  This completes our
development of the perturbation expansion, and the proof of gauge
invariance of the continuous formulation.


\subsection{Discretized form of lower-order expressions}
\label{discr}
We examine now
the discretized form of the lower-order terms for the energy and
the Sternheimer equation.
In practical calculations,
it is mandatory to use a discrete set of $k$-points
to evaluate the Brillouin-zone integrals.
However, when the focus is on $E^{(2)}$,
the discretization of the $\frac{\partial}{\partial k}$ operation
can be avoided, as the projection on the conduction
bands of the derivative of the
wavefunctions versus $k$ can be computed from
a Sternheimer equation~\cite{gonze5}. This has the disadvantage
to add a significant coding and computational step
in the whole procedure.

We choose the following symmetric finite-difference expansion for the
derivatives with respect to $k$:
\begin{eqnarray}
\label{k-discr}
\frac{\partial}{\partial k} \left.\left| u_{nk} \right.\right\rangle 
\left\langle\left. u_{nk} \right| \right.
\rightarrow \frac{1}{2\Delta k} 
\left(\vphantom{\sum_n^n}
\left.\left| u_{nk_{j+1}}\right.\right\rangle
\left\langle \left. u_{nk_{j+1}}\right| \right. 
- \left.\left| u_{nk_{j-1}}\right.\right\rangle
\left\langle\left. u_{nk_{j-1}}\right|\right. 
\right)\;,
\end{eqnarray}
where $\Delta k = k_{j+1} - k_j = (2\pi/a N_k)$. Clearly, this
expression retains the gauge invariance of the continuous form. Next,
Eq.~\ref{k-discr} is used in the derivation of explicit discretized
expressions for the energy derivatives up to the fourth-order, and 
for the Sternheimer equation up
to the second order.


\subsubsection{Energy}

From Eqs.~\ref{E2-invar} and \ref{k-discr} we obtain the discretized formula
for $E^{(2)}$:
\begin{eqnarray}
\label{E2-discr}
E^{(2)} &=& \frac{2}{N_k} \sum_{j=1}^{N_k} 
\left\{
\sum_{n=1}^N 
\left\langle u_{nk_j}^{(1)} \left|
\left( H_{k_j}^{(0)} - \varepsilon^{(0)}_{nk_j}\right) 
\right| u_{nk_j}^{(1)}\right\rangle 
\right. \nonumber \\
&&+\frac{ie}{2 \Delta k}\sum_{n,m=1}^N 
\left[
\left\langle \left. u_{n k_{j}}^{(1)}\right| \right.
\left(\vphantom{\sum_n^n}
\left| \left. u_{m k_{j+1}}^{(0)} \right\rangle \right. 
\left\langle \left. u_{m k_{j+1}}^{(0)} \right| \right. 
- \left| \left. u_{m k_{j-1}}^{(0)} \right\rangle\right. 
\left\langle \left. u_{m k_{j-1}}^{(0)}\right| \right.
\right) 
\left| \left. u_{nk_{j}}^{(0)}\right\rangle\right.
\right.\nonumber \\ 
&&-\left.\left.
\left\langle \left. u_{n k_{j}}^{(0)}\right| \right.
\left(\vphantom{\sum_n^n}
\left| \left. u_{m k_{j+1}}^{(0)} \right\rangle \right. 
\left\langle \left. u_{m k_{j+1}}^{(0)} \right| \right. 
- \left| \left. u_{m k_{j-1}}^{(0)} \right\rangle\right. 
\left\langle \left. u_{m k_{j-1}}^{(1)}\right| \right.
\right) 
\left| \left. u_{nk_{j}}^{(0)}\right\rangle\right.
\right]\right\} \;.
\end{eqnarray}

The non-variational expression for $E^{(2)}$ is written
\begin{eqnarray}
\label{E2-discr-nv}
E^{(2)} &=& -\frac{2}{N_k} \sum_{j=1}^{N_k} 
\sum_{n=1}^N 
\left\langle u_{nk_j}^{(1)} \left|
\left( H_{k_j}^{(0)} - \varepsilon^{(0)}_{nk_j}\right) 
\right| u_{nk_j}^{(1)}\right\rangle 
\nonumber \\
&=& \frac{ie}{2 \Delta k}\sum_{n,m=1}^N 
\left[
\left\langle \left. u_{n k_{j}}^{(1)}\right| \right.
\left(\vphantom{\sum_n^n}
\left| \left. u_{m k_{j+1}}^{(0)} \right\rangle \right. 
\left\langle \left. u_{m k_{j+1}}^{(0)} \right| \right. 
- \left| \left. u_{m k_{j-1}}^{(0)} \right\rangle\right. 
\left\langle \left. u_{m k_{j-1}}^{(0)}\right| \right.
\right) 
\left| \left. u_{nk_{j}}^{(0)}\right\rangle\right.
\right.\nonumber \\ 
&&-\left.
\left\langle \left. u_{n k_{j}}^{(0)}\right| \right.
\left(\vphantom{\sum_n^n}
\left| \left. u_{m k_{j+1}}^{(0)} \right\rangle \right. 
\left\langle \left. u_{m k_{j+1}}^{(0)} \right| \right. 
- \left| \left. u_{m k_{j-1}}^{(0)} \right\rangle\right. 
\left\langle \left. u_{m k_{j-1}}^{(1)}\right| \right.
\right) 
\left| \left. u_{nk_{j}}^{(0)}\right\rangle\right.
\right]\;.
\end{eqnarray}

The discretized versions of Eqs.~\ref{E3-invar} and \ref{E4-invar} are
\begin{eqnarray}
\label{E3-discr}
E^{(3)}&=&\frac{ie}{N_k\Delta k}
\sum_{n,m=1}^N  \left\langle \left. u_{n k_{j}}^{(1)}\right| \right.
\left(\vphantom{\sum_n^n}
\left| \left. u_{m k_{j+1}}^{(1)} \right\rangle \right. 
\left\langle \left. u_{m k_{j+1}}^{(0)} \right| \right. 
- \left| \left. u_{m k_{j-1}}^{(1)} \right \rangle\right. 
\left\langle \left. u_{m k_{j-1}}^{(0)}\right| \right.
\right) 
\left| \left. u_{nk_{j}}^{(0)}\right\rangle\right.
\;.\\ 
\nonumber\\
\label{E4-discr}
E^{(4)}&=& \frac{2}{N_k} \sum_{j=1}^{N_k} 
\left\{
\sum_{n=1}^N 
\left\langle u_{nk_j}^{(2)} \left|
\left( H_{k_j}^{(0)} - \varepsilon^{(0)}_{nk_j}\right) 
\right| u_{nk_j}^{(2)}\right\rangle 
\right. \nonumber \\
&&+\frac{ie}{2 \Delta k}\sum_{n,m=1}^N 
\left[
\left\langle \left. u_{n k_{j}}^{(2)}\right| \right.
\left(\vphantom{\sum_n^n}
\left| \left. u_{m k_{j+1}}^{(1)} \right\rangle \right. 
\left\langle \left. u_{m k_{j+1}}^{(0)} \right| \right. 
- \left| \left. u_{m k_{j-1}}^{(1)} \right\rangle\right. 
\left\langle \left. u_{m k_{j-1}}^{(0)}\right| \right.
\right) 
\left| \left. u_{nk_{j}}^{(0)}\right\rangle\right.
\right.\nonumber \\ 
&&-\left.\left.
\left\langle \left. u_{n k_{j}}^{(0)}\right| \right.
\left(\vphantom{\sum_n^n}
\left| \left. u_{m k_{j+1}}^{(0)} \right\rangle \right. 
\left\langle \left. u_{m k_{j+1}}^{(1)} \right| \right. 
- \left| \left. u_{m k_{j-1}}^{(0)} \right\rangle\right. 
\left\langle \left. u_{m k_{j-1}}^{(1)}\right| \right.
\right) 
\left| \left. u_{nk_{j}}^{(2)}\right\rangle\right.
\right] \vphantom{\sum_n^N}\right\} \;.
\end{eqnarray}

\subsubsection{Sternheimer equation}

The discretized expressions for the $i=1$ and $i=2$ terms of the
Sternheimer equation are given by
\begin{eqnarray}
\label{stern1-discr}
P_{ck_j}\left( H_{k_j}^{(0)} - \varepsilon^{(0)}_{nk_j}\right) P_{ck_j} 
\left.\left| u_{nk_j}^{(1)}\right. \right\rangle = 
-\frac{ie}{2 \Delta k}
P_{ck_j}\sum_{m=1}^N 
\left(\vphantom{\sum_n^N}
\left.\left| u_{m k_{j+1}}^{(0)}\right. \right\rangle
\left\langle \left. u_{m k_{j+1}}^{(0)} \right| \right. 
- \left. \left| u_{m k_{j-1}}^{(0)} \right. \right\rangle
\left\langle \left. u_{mk_{j-1}}^{(0)} \right| \right. 
\right) 
\left.\left| u_{nk_{j}}^{(0)}\right. \right\rangle\;,\\ 
\nonumber\\
\label{stern2-discr}
P_{ck_j}\left( H_{k_j}^{(0)} - \varepsilon^{(0)}_{nk_j}\right) P_{ck_j} 
\left.\left| u_{nk_j}^{(2)}\right. \right\rangle = 
-\frac{ie}{2 \Delta k}
P_{ck_j}\sum_{m=1}^N 
\left(\vphantom{\sum_n^N}
\left.\left| u_{m k_{j+1}}^{(1)}\right. \right\rangle
\left\langle \left. u_{m k_{j+1}}^{(0)} \right| \right. 
- \left. \left| u_{m k_{j-1}}^{(0)} \right. \right\rangle
\left\langle \left. u_{mk_{j-1}}^{(1)} \right| \right. 
\right) 
\left.\left| u_{nk_{j}}^{(0)}\right. \right\rangle\;. 
\end{eqnarray}
The required gauge invariance of the expansion terms is preserved in
the discretized expressions. Note that the solutions at a given
k-point are now coupled to the first-neighbor k-points in the
reciprocal-space grid.


\section{Perturbation theory applied to the discretized polarization}
\label{sec:pt-P}

\subsection{General perturbation expansion}
\label{expan-P} 

Let us consider now the perturbation treatment of the problem on the
basis of the energy functional given in Eq.~\ref{E-def-P}, where the
polarization is written in a discretized form. This formulation can be
viewed as the reciprocal-space analog of the NV~\cite{nv} real-space
functional. In this approach the gauge-invariance of the energy is
guaranteed by the fact that Eq.~\ref{P-discr} is itself gauge
invariant.~\cite{resta1,ksv,vks}

We seek a minimum for Eq.~\ref{E-def-P} with respect to the occupied
orbitals $\left\{ u_{nk_j}\right\}$, under the constraints
$\left\langle u_{nk_j} \left| u_{mk_j}\right\rangle \right. =
\delta_{mn}$. Lagrange multipliers are introduced to write the
unconstrained functional
\begin{eqnarray}
\label{E1-P}
{\cal F}\left[\left\{u_{nk_j}\right\};{\cal E}\right] 
&=& \frac{2}{N_k} \sum_{j=1}^{N_k} \left[ \sum_{n,m=1}^N  
\left\langle u_{mk_j} \left| H^{(0)}_{k_j} \right| u_{nk_j} 
\right\rangle~\delta_{mn} - \left( \left\langle u_{nk_j} 
\left| u_{mk_j}\right\rangle\right. 
- \delta_{mn} \right)\Lambda_{mn}(k_j)\right. \nonumber \\
&-& \left. \left( \frac{e{\cal E}}{\Delta k}\right) 
{\rm Im} \left\{ \ln \det \left[ S_{nm}(k_j,k_{j+1}) \right] 
\right\}\vphantom{\sum_n^N}\right]\;.
\end{eqnarray}

In Appendix~\ref{app:S-series}, we prove the following result:
\begin{eqnarray}
\frac{\delta \sum_j {\rm Im}\left\{\ln \det \left[
S_{m^\prime m}(k_j,k_{j+1}) \right] \right\}} {\delta u_{nk_j}^\ast} =
-\frac{i}{2} \sum_{m=1}^N \left[ 
\left.\left| u_{mk_{j+1}}\right. \right\rangle
S^{-1}_{mn}(k_j,k_{j+1}) 
- \left.\left| u_{mk_{j-1}}\right. \right\rangle
S^{-1}_{mn}(k_j,k_{j-1}) \right]\;,
\end{eqnarray}
which can be used to derive the Euler-Lagrange
equation from Eq.~\ref{E1-P}, as follows:
\begin{eqnarray}
\label{stern0-P}
\frac{\delta {\cal F}}{\delta u^\ast_{nk_j}} &=& \frac{2}{N_k} 
\left\{
H_{k_j}^{(0)} \left.\left| u_{nk_j}\right.\right\rangle - 
\sum_{m=1}^N 
\left.\left| u_{mk_j} \right.\right\rangle
\Lambda_{mn}(k_j) 
\right. \nonumber \\ 
&+& \left. 
\left(\frac{i e{\cal E}}{2\Delta k}\right)
\sum_{m=1}^N 
\left[ 
\left. \left| u_{mk_{j+1}} \right. \right\rangle
S^{-1}_{mn}(k_j,k_{j+1}) 
- \left.\left| u_{mk_{j-1}} \right. \right\rangle
S^{-1}_{mn}(k_j,k_{j-1}) 
\right] 
\vphantom{\sum_n^N}\right\} = 0\;.
\end{eqnarray}

Below, we consider the perturbation expansions of the Lagrange
multipliers, the energy, and the Sternheimer equation. The expansion
of the orthonormalization condition was already developed in the
previous section, and remains unaltered in the present case.


\subsubsection{Lagrange multipliers}
We multiply Eq.~\ref{stern0-P} on the left by $u^\ast_{mk_j}$ to write the
Lagrange multipliers
\begin{eqnarray}
\label{LM-P}
\Lambda_{mn}(k_j) = \left\langle u_{mk_j} 
\left| H_{k_j}^{(0)} \right| u_{nk_j} \right\rangle\;.
\end{eqnarray}
It can be readily seen that the terms involving the overlap matrix
cancel out, since
\begin{eqnarray}
&&\sum_l \left[ 
\left\langle u_{mk_j} \left| u_{l k_{j+1}} \right\rangle \right. 
S^{-1}_{l n}(k_j,k_{j+1}) 
- \left\langle u_{m k_j} \left| u_{l k_{j-1}} \right\rangle \right. 
S^{-1}_{l n}(k_j,k_{j-1}) \right]  \nonumber \\
&~& ~= \sum_l
\left[ S_{ml}(k_j,k_{j+1}) S^{-1}_{l n}(k_j,k_{j+1}) - 
S_{ml}(k_j,k_{j-1}) S^{-1}_{l n}(k_j,k_{j-1}) \right]  
= 0\;.
\end{eqnarray}

The perturbation expansion of Eq.~\ref{LM-P} takes the
simple form
\begin{eqnarray}
\label{LM1-P}
\Lambda_{mn}^{(i)}(k_j) = \sum_{j=0}^i 
\left\langle u_{mk_j}^{(i-j)} \left| 
H_{k_j}^{(0)} \right| u_{nk_j}^{(j)} \right\rangle\;.
\end{eqnarray}


\subsubsection{Energy}
\label{Energy-P}

In order to write the perturbation expansion of Eq.~\ref{E1-P}, we
need the expansion of the polarization on the basis of the $2n+1$
theorem. The variation-perturbation framework allows us to focus on
the part of $E_{pol}^{(2i)}$ or $E_{pol}^{(2i+1)}$ that comes from
variation of the wavefunctions up to order $i$ only.  For these
quantities, we introduce the notation $E_{pol}^{(2i,i)}$ or
$E_{pol}^{(2i+1,i)}$. The even terms are written
\begin{eqnarray}
\label{P-even-discr}
E_{pol}^{(2i,i)} &=& -\left[{\cal E} P\left(\sum_{j=0}^i
{\cal E}^j u_{nk_j}^{(j)} \right) \right] ^{(2i)} = \left.
-\frac{1}{(2i)!} \frac{\partial^{2i}}{\partial {\cal
E}^{2i}} \left[{\cal E} P\left(\sum_{j=0}^i {\cal E}^j
u_{nk_j}^{(j)} \right) \right] \right|_{{\cal E}=0} \nonumber\\ 
&=& -\left[P\left(\sum_{j=0}^i {\cal E}^j u_{nk_j}^{(j)}
\right) \right]^{(2i-1)}\;.
\end{eqnarray}

By the same token, for the odd terms we obtain
\begin{eqnarray}
\label{P-odd-discr}
E_{pol}^{(2i+1,i)} = -\left[{\cal E} P\left(\sum_{j=0}^i
{\cal E}^j u_{nk_j}^{(j)} \right) \right]^{(2i+1)}
=-\left[P\left(\sum_{j=0}^i {\cal E}^j u_{nk_j}^{(j)}
\right) \right]^{(2i)}\;.
\end{eqnarray}

In these expressions, the $2n+1$-theorem implies that only the
contributions of order $\leq i$ from the perturbed wave functions will
appear, when we consider the contribution of the polarization term to
the total-energy derivatives. More explicit formulas for computing
these polarization derivatives will be given below.

With these results, we can expand Eq.~\ref{E1-P}. From
Eqs.~\ref{E-pert} and \ref{P-even-discr} we obtain the even terms
\begin{eqnarray}
\label{E-even-P}  
E^{(2i)} &=& 
E_{pol}^{(2i,i)} + \frac{2}{N_k} \sum_{j=1}^{N_k}
\sum_{m,n=1}^N 
\left[ 
\vphantom{\sum_n^N}\left\langle u_{mk_j}^{(i)} 
\left| H_{k_j}^{(0)} \right|
u_{nk_j}^{(i)}\right\rangle~\delta_{mn} \right. \nonumber\\
&&\left. 
-\sum_{l^\prime,l^{\prime\prime}=1}^i
\sum_{l=0}^{i-1} \delta(2i-l-l^\prime-l^{\prime\prime})~
\Lambda_{mn}^{(l)}(k_j) 
\left\langle u_{nk_j}^{(l^\prime)} \left| 
u_{mk_j}^{(l^{\prime\prime)}}\right.\right\rangle 
\right]\;,
\end{eqnarray}
while from Eqs.~\ref{E-pert} and \ref{P-odd-discr}, the odd terms are written
\begin{eqnarray}
\label{E-odd-P}
E^{(2i+1)} = E_{pol}^{(2i+1,i)} +  \frac{2}{N_k} \sum_{j=1}^{N_k} \sum_{m,n=1}^N 
\sum_{l^\prime,l^{\prime\prime}=1}^i
\sum_{l=1}^{i} \delta(2i+1-l-l^\prime-l^{\prime\prime})~
\Lambda_{mn}^{(l)}(k_j) 
\left\langle u_{nk_j}^{(l^\prime)} \left| u_{mk_j}^{(l^{\prime\prime)}}
\right\rangle\right.\;.
\end{eqnarray}
%


\subsubsection{Sternheimer equation}
The perturbation expansion of Eq.~\ref{stern0-P} yields the
Sternheimer equation
\begin{eqnarray}
\label{stern.i-P}
&&P_{ck_j} \left( H^{(0)}_{k_j} - \varepsilon^{(0)}_{nk_j}\right) P_{ck_j} 
\left.\left| u_{nk_j}^{(i)} \right.\right\rangle =  
\sum_{m=1}^{N} 
\left\{ 
\sum_{l=1}^{i-1} P_{c {k_j}} 
\left.\left| u_{m {k_j}}^{(i-l)}\right.\right\rangle 
\Lambda_{mn}^{(l)}(k_j)
\right. \nonumber \\  
&~& \left. 
- \frac{i e}{2\Delta k} \sum_{l=0}^{i-1}  P_{c {k_j}} 
\left[ 
\vphantom{\sum_n^N}
\left.\left| u_{mk_{j+1}}^{(i-l-1)} \right.\right\rangle 
S_{mn}^{-1(l)}(k_j,k_{j+1}) 
- \left.\left| u_{mk_{j-1}}^{(i-l-1)} \right.\right\rangle 
S_{mn}^{-1(l)}(k_j,k_{j-1}) 
\right]
\right\} = 0\;.
\end{eqnarray}

The conduction-band projector $P_{ck}$ in Eqs.~\ref{stern.i} and
\ref{stern.i-P} appears due the fact that the Lagrange multipliers in
these two equations may be different, due to $k$-gauge freedom. Thus,
only conduction-band contributions can be identified while comparing
the two formulations.

By comparing Eqs.~\ref{stern.i} and
\ref{stern.i-P}, we see that in the
present formulation the term $ie~\!P_{ck}\frac{\partial}{\partial k}
\left.\left| u_{nk} \right. \right\rangle$ is approximated by the 
finite-difference formula
\begin{eqnarray}
\label{k-P}
D(\Delta k)=
\frac{i e}{2\Delta k} \sum_{m=1}^N P_{ck_j} 
\left[ 
\vphantom{\sum_n^N}
\left.\left| u_{mk_{j+1}}\right.\right\rangle 
S^{-1}_{mn}(k_j,k_{j+1}) 
- \left.\left| u_{mk_{j-1}} \right.\right\rangle
S^{-1}_{mn}(k_j,k_{j-1}) 
\right]\;.
\end{eqnarray} 

The theory of finite-difference approximations to derivatives
could now be applied to Eq.~\ref{k-P}, as a function of $\Delta k$.
Note that this expression is invariant under $\Delta k \rightarrow -\Delta k$,
as it induces simultaneous exchange of $k_{j+1}$ and $k_{j-1}$.
Thus
\begin{eqnarray}
D(\Delta k)=D(0)+{\cal O}(\Delta k)^2.
\end{eqnarray}
One can now define
\begin{eqnarray}
\label{2k-P}
D(2 \Delta k)=
\frac{i e}{2.2\Delta k} \sum_{m=1}^N P_{ck_j}
\left[
\vphantom{\sum_n^N}
\left.\left| u_{mk_{j+2}}\right.\right\rangle
S^{-1}_{mn}(k_j,k_{j+2})
- \left.\left| u_{mk_{j-2}} \right.\right\rangle
S^{-1}_{mn}(k_j,k_{j-2})
\right]\;,
\end{eqnarray}
giving a higher-order approximation of $D(0)$ as
\begin{eqnarray}
D^{higher-order}=\left[ 4 D(\Delta k) - D(2 \Delta k)\right]/3=D(0)+{\cal O}(\Delta k)^4\;.
\end{eqnarray}

This improved expression also derives from a total energy functional.
Instead of Eq.~\ref{E-def-P}, one must start from
\begin{eqnarray}
\label{Ehigh-def-P}
E^{higher-order}\left[\left\{u_{nk_j}\right\};{\cal E}\right] &=&
\frac{2}{N_k}
\sum_{n=1}^N \sum_{j=1}^{N_k} \left\langle u_{nk_j}
\left| H^{(0)}_{k_j} \right| u_{nk_j} \right\rangle
\nonumber \\
&-&
\frac{2e{\cal E}}{N_k\Delta k} \sum_{j=1}^{N_k} {\rm Im} \left\{
\frac{4}{3} \ln \det \left[ S_{mn}(k_j,k_{j+1}) \right]
- \frac{1}{6} \ln \det \left[ S_{mn}(k_j,k_{j+2}) \right]
\right\}\;.
\end{eqnarray}
Despite its interest, we will not explore
this topics further in the present paper.


\subsection{Lower-order expressions}
\label{lower-P}
In Appendix~\ref{app:S-series}, we derive the Taylor expansion of
Eq.~\ref{P-discr} for the polarization, which allows us to obtain
explicit expressions for Eqs.~\ref{P-even-discr} and
\ref{P-odd-discr}. Here, we look at the lower-order expressions for
the energy and the Sternheimer equation.


\subsubsection{Energy}
From Appendix~\ref{app:S-series}, the second-order polarization term is
given by
\begin{eqnarray}
\label{Epol-2}
E_{pol}^{(2,1)} &=& - \left[ P\left( \left\{ u_{nk_j}^{(0)} + {\cal E}
u_{nk_j}^{(1)} \right\} \right) \right]^{(1)}
= -\frac{2 e}{N_k\Delta k} \sum_{j=1}^{N_k} 
{\rm Im} \left\{ {\rm Tr}\left[ 
S^{(1)}(k_j,k_{j+1}) Q(k_j,k_{j+1}) 
\right] \right\}\nonumber\\
&=& -\frac{2 e}{N_k\Delta k} \sum_{j=1}^{N_k} 
{\rm Im} \left\{ \sum_{m,n=1}^N \left[
\left\langle u_{nk_j}^{(1)} 
\left| u_{mk_{j+1}}^{(0)} \right\rangle\right. + 
\left\langle u_{nk_j}^{(0)} 
\left| u_{mk_{j+1}}^{(1)} \right\rangle \right.
\right] Q_{mn}(k_j,k_{j+1})\right\} \;,
\end{eqnarray}
where $Q(k_j,k_{j+1})$, obeying
\begin{equation}
\sum_l Q_{ml}(k_j,k_{j+1}) S^{(0)}_{l n}(k_j,k_{j+1}) = \delta_{mn}\,,
\end{equation}
is the inverse of the
zeroth-order overlap matrix $S^{(0)}_{nm}(k_j,k_{j+1}) = \left\langle
u_{nk_j}^{(0)} \left| u_{mk_{j+1}}^{(0)}\right\rangle\right.\;.$

The second-order expression for the energy is then given by
the $i=1$ term in Eq.~\ref{E-even-P}
\begin{eqnarray}
\label{E2-P}
E^{(2)} &=& \frac{2}{N_k} \sum_{j=1}^{N_k} \sum_{n=1}^N 
\left\langle u_{nk_j}^{(1)} \left| 
\left( H_{k_j}^{(0)} - \varepsilon_{nk_j}^{(0)} \right)
\right| u_{nk_j}^{(1)} \right\rangle 
\nonumber\\
&&-\frac{e}{\Delta k} \sum_{j=1}^{N_k} {\rm Im} 
\left\{ 
\sum_{m,n=1}^N 
\left(
\left\langle u_{nk_j}^{(1)} \left| u_{mk_{j+1}}^{(0)} \right\rangle\right. 
+ \left\langle u_{nk_j}^{(0)} \left| u_{mk_{j+1}}^{(1)} \right\rangle\right.
\right) 
Q_{mn}(k_j,k_{j+1}) 
\right\}\;.
\end{eqnarray}

The third-order derivative of the energy is given by by the $i=1$ in
Eq.~\ref{E-odd-P}. The first-order contribution to the Lagrange
multipliers vanishes, since from Eqs.~\ref{u-ortho} and \ref{LM1-P} we
have $\Lambda_{mn}^{(1)}(k_j) = \left\langle u_{mk_j}^{(1)} \left|
H_{k_j}^{(0)} \right| u_{nk_j}^{(0)} \right\rangle + \left\langle
u_{mk_j}^{(0)} \left| H_{k_j}^{(0)} \right| u_{nk_j}^{(1)}
\right\rangle = \varepsilon_{nk_j}^{(0)} \left\langle u_{mk_j}^{(1)}
\left| u_{nk_j}^{(0)} \right\rangle \right. + \varepsilon_{mk_j}^{(0)}
\left\langle u_{mk_j}^{(0)} \left| u_{nk_j}^{(1)}
\right\rangle\right. = 0$, which leads to $E^{(3)} =
E_{pol}^{(3,1)}$.  From the results in Appendix~\ref{app:S-series}
\begin{eqnarray}
\label{E3-P}
E^{(3)} &=& - 
\left[ 
P\left(
\left\{u_{nk_j}^{(0)} + F~u_{nk_j}^{(1)} \right\} 
\right) 
\right]^{(2)} \nonumber\\
&=& -\frac{e}{N_k\Delta k} \sum_{j=1}^{N_k} {\rm Im} 
\left\{ 
{\rm Tr}
\left[
2S^{(2)}(k_j,k_{j+1})~Q(k_j,k_{j+1})
\right.\right.\nonumber\\ 
&&- \left.\left.
S^{(1)}\left( k_j,k_{j+1}\right) Q\left( k_j,k_{j+1}\right)
S^{(1)}\left( k_j,k_{j+1}\right) Q\left( k_j,k_{j+1}\right)
\right]\right\}\nonumber\\ 
&=&-\frac{e}{N_k\Delta k}\sum_{j=1}^{N_k} {\rm Im}
\left\{ 
\sum_{m,n=1}^N 
2\left\langle u_{nk_j}^{(1)} \left| u_{mk_{j+1}}^{(1)}\right.\right\rangle 
Q_{mn}\left( k_j,k_{j+1}\right) 
\right.\nonumber\\
&&- \sum_{m,n,l,l^\prime=1}^N 
\left[ 
\left\langle u_{mk_j}^{(1)}\left| u_{n k_{j+1}}^{(0)}\right\rangle\right. + 
\left\langle u_{mk_j}^{(0)}\left| u_{n k_{j+1}}^{(1)}\right\rangle\right. 
\right]
Q_{nl}\left( k_j,k_{j+1}\right) \nonumber \\
&& \left. \left[ 
\left\langle u_{l k}^{(1)} \left| 
u_{l^\prime k_{j+1}}^{(0)}\right.\right\rangle + 
\left\langle u_{l k}^{(0)} \left| 
u_{l^\prime k_{j+1}}^{(1)}\right.\right\rangle 
\right]
Q_{l^\prime m}\left( k_j,k_{j+1}\right)
\vphantom{\sum_{m}^N}\right\}\;.
\end{eqnarray}

For the fourth- and higher-order energy derivatives, the expansion
yields very involved expressions. We end this section by considering
the fourth-order term for the energy in a more compact notation [we
drop the $\left( k_j,k_{j+1}\right)$ matrix arguments]:
\begin{eqnarray}
\label{Epol-4}
E_{pol}^{(4,2)} &=& - 
\left[ 
P\left(
\left\{
u_{nk_j}^{(0)} + F~u_{nk_j}^{(1)} + F^2~u_{nk_j}^{(2)}
\right\} 
\right) 
\right]^{(3)} \nonumber\\
&=& -\frac{2 e}{3 N_k\Delta k} \sum_{j=1}^{N_k} {\rm Im} 
\left\{ 
{\rm Tr}
\left[ 
3~\!S^{(3)} Q - 3~\! S^{(2)}~Q~S^{(1)}~Q + 
S^{(1)}~Q~S^{(1)}~Q~S^{(1)}~Q 
\right] \right\}\nonumber\\
&=&-\frac{2 e}{3 N_k\Delta k} \sum_{j=1}^{N_k} {\rm Im}
\left\{ 
3\sum_{m,n=1}^N 
\left[
\left\langle u_{nk_j}^{(1)} \left| u_{mk_{j+1}}^{(2)}\right.\right\rangle 
+\left\langle u_{nk_j}^{(2)} \left| u_{mk_{j+1}}^{(1)}\right.\right\rangle
\right]
Q_{mn}
\right. \nonumber\\ 
&-& 3 \sum_{m,n=1}^N 
\left[ 
\left\langle u_{nk_j}^{(0)}\left| u_{n k_{j+1}}^{(2)}\right.\right\rangle + 
\left\langle u_{nk_j}^{(1)}\left| u_{m k_{j+1}}^{(1)}\right.\right\rangle + 
\left\langle u_{nk_j}^{(2)}\left| u_{m k_{j+1}}^{(0)}\right.\right\rangle 
\right]
\left( Q S^{(1)} Q\right)_{mn}
\nonumber\\
&+&\left. S^{(1)}~Q~S^{(1)}~Q~S^{(1)}~Q\vphantom{\sum_{m}^N}
\right\}
\end{eqnarray}
In this expression, we write explicitly only the terms containing
$u_{nk_j}^{(2)}$, which will determine the second-order term of the
Sternheimer equation. The corresponding fourth-order energy is given
by
\begin{eqnarray}
\label{E4-P}
E^{(4)} = \frac{2}{N_k} \sum_{k=1}^{N_k} \sum_{n=1}^N \left\langle
u_{nk_j}^{(2)} \left| \left( H_{k_j}^{(0)} - \varepsilon_{nk_j}^{(0)} \right)
\right| u_{nk_j}^{(2)} \right\rangle + E_{pol}^{(4,2)}\;.
\end{eqnarray}

\subsubsection{Sternheimer equation}

From Eq.~\ref{stern.i-P}, the first-order term for the Sternheimer
equation reads
\begin{eqnarray}
\label{stern1-P}
&&P_{ck_j} \left( H^{(0)}_{k_j} - \varepsilon^{(0)}_{nk_j}\right) P_{ck_j} 
\left.\left| u_{nk_j}^{(1)} \right. \right\rangle =  
- \frac{i e}{2\Delta k} P_{c k} \sum_{m=1}^N 
\left[ 
\left.\left| u_{mk_{j+1}}^{(0)} \right.\right\rangle 
Q_{mn}\left(k_j,k_{j+1}\right) 
- \left.\left| u_{mk_{j-1}}^{(0)}\right. \right\rangle 
Q_{mn}\left(k_j,k_{j-1}\right) 
\right]\;;\nonumber\\
\end{eqnarray}
while the second-order derivative is written
\begin{eqnarray}
\label{stern2-P}
&&P_{ck_j} \left( H^{(0)}_{k_j} - \varepsilon^{(0)}_{nk_j}\right) P_{ck_j} 
\left.\left| u_{nk_j}^{(2)} \right. \right\rangle =  
- \frac{i e}{2\Delta k} P_{c k} \left\{\sum_{m=1}^N 
\left[ 
\left.\left| u_{mk_{j+1}}^{(1)} \right. \right\rangle 
Q_{mn}\left( k_j,k_{j+1}\right) 
- \left.\left| u_{mk_{j-1}}^{(1)} \right.\right\rangle 
Q_{mn}\left( k_j,k_{j-1}\right) 
\right] \right. \nonumber\\
&&-\sum_{m,l,l^\prime=1}^N 
\left[ 
\left.\left| u_{mk_{j+1}}^{(0)} \right.\right\rangle 
Q_{m l}\left( k_j,k_{j+1}\right) 
S^{(1)}_{l,l^\prime}\left( k_j,k_{j+1}\right) 
Q_{l^\prime n}\left( k_j,k_{j+1}\right)
\right.\nonumber\\
&&\left.\left. -\left.\left| u_{mk_{j-1}}^{(0)} \right.\right\rangle 
Q_{m l}\left( k_j,k_{j-1}\right)
S^{(1)}_{l,l^\prime}\left( k_j,k_{j-1}\right) 
Q_{l^\prime n}\left( k_j,k_{j-1}\right)
\right]
\vphantom{\sum_n^N}
\right\}\;.
\end{eqnarray}
These two expressions can also be consistently obtained by taking the
conduction-band projection of the corresponding terms in the expansion of
$\delta E^{(2i)}/\delta u^{\ast(i)}_{nk_j} = 0$.


\section{Numerical tests}
\label{sec:1d-model}

In this section, we illustrate the present theory by applying it the
two-band 1D-TB Hamiltonian introduced in Sec.~\ref{sec:efp}.  For this
model, exact analytical expressions can be written for the continuous
formulation. Our purpose in this simple application is to demonstrate
consistency between the continuous and the discretized versions of the
theory, and also to make a preliminary assessment of the convergence
of the energy derivatives obtained in the two discretized
formulations, with respect to $k$-point sampling in the Brillouin
zone.  Appendix~\ref{app:1dmodel} contains the detailed derivations of
the results presented here. As such, it presents a step-by-step
example of the use of the formalism developed in the present paper.

\subsection{Response to a homogeneous electric field}
In the continuous formulation, the first-order change of the valence
state is obtained from the corresponding term in the Sternheimer
equation. By setting $i=1$ in Eq.~\ref{stern.i}, this is given by
\begin{eqnarray}
\label{1d-str1}
P_{ck} \left( H_k^{(0)} -
\varepsilon_{vk}^{(0)} \right) P_{ck} 
\left.\left| u_{vk}^{(1)} \right.\right\rangle = 
- P_{ck} i \frac{\partial}{\partial k}
\left. \left| u_{vk}^{(0)} \right.\right\rangle\;. 
\end{eqnarray}
The solution to this equation is given in Appendix~\ref{app:1dmodel}.

Having solved Eq.~\ref{1d-str1}, $E^{(2)}$ can then be obtained from
the simplest non-variational expression Eq.~\ref{E2-invar-nv} as
\begin{eqnarray}
\label{1d-E2}
E^{(2)} = - {1 \over \pi} \int_0^{2\pi} dk {1 \over \Delta\varepsilon_k} 
\left({\partial \Theta_k \over \partial k}\right)^2\;,
\end{eqnarray}
where $\Delta\varepsilon_k = \varepsilon_{ck}^{(0)} -
\varepsilon_{vk}^{(0)}$, and $\partial \Theta_k / \partial k$ is
obtained from Eqs.~\ref{1d-eigval} and \ref{theta} (see
Eq.~\ref{dthetadk}).

The second-order change of the valence state is obtained from the
$i=2$ term in Eq.~\ref{stern.i} and from Eq.~\ref{Pv-u2}.  In our 1D
model, the Sternheimer equation for the conduction-band projection
reads
\begin{eqnarray}
\label{1d-str2}
P_{ck} \left( H_k^{(0)} - \varepsilon_{vk}^{(0)} \right) P_{ck} 
\left.\left| u_{vk}^{(2)} \right.\right\rangle = - P_{ck} 
\left[ i \frac{\partial}{\partial k}
\left.\left| u_{vk}^{(1)}\right. \right\rangle 
- \Lambda_k^{(1)} \left.\left| u_{vk}^{(1)}
\right\rangle \right.\right] \;.
\end{eqnarray}

Using the solution for $\left.\left| u_{vk}^{(2)}
\right.\right\rangle$ given in Appendix~\ref{app:1dmodel}, we can now
arrive at the expression for the fourth-order energy. Applying
Eq.~\ref{E4-invar}, we write
\begin{eqnarray}
\label{1d-E4}
E^{(4)} = \frac{1}{\pi} \int_0^{2\pi} dk \left[ 
\left( \frac{1}{\Delta \varepsilon_k}\right)^3 
\left( {\partial \Theta_k \over \partial k} \right)^4
- \left( {1 \over \Delta \varepsilon_k}\right) 
\left\{ {\partial \over \partial k} \left( {1 \over \Delta\varepsilon_k} 
{\partial \Theta_k \over \partial k} \right) \right\}^2 \right]\;.
\end{eqnarray}  
(Expressions for $E^{(2)}$ and $E^{(4)}$ in terms of elliptic
integrals of the second kind are given in Appendix~\ref{app:1dmodel}.)

Turning now to the discretized expressions, using the non-variational
Eq.~\ref{E2-invar-nv} 
we compute the DAPE expression for the second-order energy:
\begin{eqnarray}
\label{1d-E2-discr}
E^{(2)} = - {2 \over N_k} \sum_{j=1}^{N_k} {1 \over \Delta
\varepsilon_{k_j}} \left[ {1 \over 2\Delta k} \sin\left(\Theta_{j+1} -
\Theta_{j-1} \right) \cos\left(2\Theta_j -\Theta_{j+1} - \Theta_{j-1}
\right) \right]^2\;.
\end{eqnarray}
The PEAD second-order energy expression takes a different form: 
\begin{eqnarray}
\label{P1d-E2}
E^{(2)} = - {2 \over N_k} \sum_{k=1}^{N_k} {1 \over \Delta \varepsilon_j}
\left\{ {1 \over 2\Delta k} \left[\vphantom{\frac{}{}}
\tan\left(\Theta_{j+1} - \Theta_j \right)
-\tan\left(\Theta_{j-1} - \Theta_j \right)
\right] \right\}^2\;.
\end{eqnarray}

With the results in Appendix~\ref{app:1dmodel}, we have all the
ingredients to write analytical expressions for the discretized
versions of $E^{(4)}$, but these are quite cumbersome and
will not be reproduced here. The numerical results obtained using
these expressions are discussed below.


\subsection{Numerical results}
\label{numerical}
In order to test the consistency between the three formulations, we
checked that by sufficiently increasing the number of $k$-points used
in the evaluation of the discretized energy derivatives, we obtain an
agreement between continuous and discretized expressions within
stringent degrees of accuracy.  For example, for $t=1$, 80 $k$-points
in the full Brillouin zone are needed for the three expressions for
$E^{(2)}$ (Eqs.~\ref{1d-E2}, ~\ref{1d-E2-discr}, and ~\ref{P1d-E2}) to
agree within $\sim 1$\%, while 240 $k$-points are needed to get the
same level of agreement for the $E^{(4)}$ expressions.

By decreasing the number of $k$-points, thus worsening the level of
accuracy of the discretized expressions, the differences between them
become more apparent. In Fig.~\ref{e2fig}, we show the quantity
$\left[ E^{(2)}_{discr.} - E^{(2)}_{exact}\right]/E^{(2)}_{exact}$
giving the percentual error in the evaluation of $E^{(2)}$ for the two
discretized formulations, using 20 $k$-points, with the hoping
parameter varying over the [0,1] interval. In Fig.~\ref{e4fig} the
corresponding quantity for $E^{(4)}$, $\left[ E^{(4)}_{discr} -
E^{(4)}_{exact}\right]/E^{(4)}_{exact}$, is shown for a sampling of 80
$k$-points.

It is clear from Figs.~\ref{e2fig} and \ref{e4fig} that the energy
derivatives obtained from the PEAD converge faster with respect to the
number of $k$-points, at least for the present 1D model.  However, it
must be borne in mind that this formulation involves the calculation
of the inverse of the zero-field overlap matrix, and in practical
applications the additional cost associated with this operation could
offset the gain in $k$-point convergence, specially when the two
formulations are applied to systems with large numbers of atoms in the
unit cell. This point remains to be further addressed when the theory
is applied in the context of realistic tight-binding and {\it ab
initio} calculations.

We also computed the norm of the first- and second-order wavefunction
derivatives $\left[ \left\langle u_{vk}^{(i)} \left| u_{vk}^{(i)}
\right\rangle\right. \right]^{1/2}\;$ as a function of $k$, for the
continuous solutions and the two discretized forms, with a value of
$t=1$ for the hoping parameter and samplings of 20 and 80
$k$-points. As expected from the above results for $E^{(2)}$ and
$E^{(4)}$, we observe that the wavefunctions in the PEAD are better
approximations to the exact ones from the continuous formulation.


\section{Summary}
\label{sum}
The goal of this work was to obtain second- and higher-order derivatives
of the total energy of periodic insulators, with respect to an applied
homogeneous electric field. Related physical properties are the linear
and non-linear dielectric susceptibilities (connected to linear and
non-linear optical constants).

Although a variation-perturbation framework had been formulated
earlier for the computation of derivatives of the total energy with
respect to many different perturbations, several formal and technical
difficulties must be addressed when considering the specific case of a
homogeneous-electric-field perturbation.

At the level of the electric-field-dependent energy functional, we
proposed a basic expression, Eq.~\ref{E-def}, that we argue to be valid
in the space of states possessing a periodic density. It is directly
linked to the modern theory of polarization, proposed by King-Smith
and Vanderbilt nearly a decade ago, and allows to recover easily the Berry
phase polarization formula, central to this theory.

Unfortunately, when the polarization is varied, this energy functional
leads to local minima with a basin of attraction of {\it infinitesimal}
extent.  A regularization procedure, based on a $k$-point
discretization of the reciprocal-space integrals, must be used in
order to lead to a finite size basin. This is the reciprocal-space
analog of the real-space cut-off introduced by Nunes and Vanderbilt in
their treatment of polarized Wannier functions.

Having thus defined a suitable energy functional, that depends on the
applied homogeneous electric field, we were allowed to proceed with the
application of the variation-perturbation machinery.  Interestingly,
the derivation of the canonical formulas at all orders of perturbation
can be done either on the basis of the energy functional already
regularized, or on the basis of the unregularized one, followed by
regularization at each order. The formulas derived in the two cases
differ from each other. Working with the regularized energy functional
gives more cumbersome expressions, however perfectly consistent with
an energy functional, while the a posteriori application of the
regularization at each order is not consistent with the regularized
energy functional. The two procedures will tend to the same limit when the
discretization is refined further and further.  The expression for the
third-order derivative of the total energy, previously proposed by
Mauri and Dal Corso, is recovered, as an instance of the "a
posteriori" application of the regularization technique.

We applied this formalism to a model one-dimensional two-band
Hamiltonian, showing explicitly the pathology of a non-regularized
energy functional and its cure, as well as the differences related to
the order in which the perturbation expansion and the discretization
procedure are applied. The two discretized formulations are shown to
agree in the continuum limit, although the ``perturbation
expansion after discretization'' (PEAD) formulation seems to be closer
to the exact answer than the ``discretization after perturbation
expansion'' (DAPE) formulation, for an equivalent grid.


\appendix

\section{}
\label{app:gauge}

In this appendix, we describe the algebraic manipulations needed to
transform Eqs~\ref{E-even}, \ref{E-odd}, and \ref{stern.i} into their
gauge-invariant forms presented in Secs.~\ref{pt-cont-energy} and \ref{pt-cont-stern}. We will use the following
result:
\begin{eqnarray}
\label{a1}
\left\langle x \left| \frac{\partial}{\partial k} \right| y \right\rangle 
\left.\left| z \right.\right\rangle =
\left\langle x \left| y \right.\right\rangle \frac{\partial}{\partial k}
\left. \left| z \right.\right\rangle + 
\left. \left| z \right.\right\rangle  
\frac{\partial}{\partial k} 
\left\langle x \left| y \right. \right\rangle 
- \left(
\frac{\partial}{\partial k}\left. \left| z \right. \right\rangle
\left\langle \left. x \right|\right\rangle 
\right) \left. \left| y \right.\right\rangle\;.
\end{eqnarray}

Furthermore, from Eq.~\ref{ortho-pert} we have
\begin{eqnarray}
\label{a2}
\sum_{l=0}^{i} \left\langle u_{mk}^{(l)} \left| 
u_{nk}^{(i-l)} \right.\right\rangle = 0 ~\; ; \;~i\geq 1\;.
\end{eqnarray}

With the help of Eq.~\ref{a1} and $\delta_{mn} = \left\langle
u_{mk}^{(0)} \left| u_{nk}^{(0)} \right.\right\rangle$ , it is
straightforward to show that
\begin{eqnarray}
\label{a3}
\left[\frac{\partial}{\partial k}\delta_{mn} - \left\langle u_{mk}^{(0)}
\left| \frac{\partial}{\partial_k} \right| 
u_{n k}^{(0)} \right\rangle \right]
\left.\left| u_{m k}^{(i)} \right.\right\rangle = 
\left( 
\frac{\partial}{\partial k} 
\left.\left| u_{mk}^{(i)} \right.\right\rangle 
\left\langle \left. u_{mk}^{(0)} \right|\right. 
\right) 
\left| \left. u_{nk}^{(0)} \right\rangle\right.\;.
\end{eqnarray}
We recall that the notation ``$\left( \partial_k \left| \left. u
\right.\right\rangle \left\langle \left. u \right|\right. \right)$''
indicates that $\partial_k$ acts only on the expression in
parenthesis.

Now, for reference we repeat Eqs.~\ref{E-even} and \ref{E-odd} here:
\begin{eqnarray}
\label{a4}  
E^{(2i)} &=& \frac{a}{\pi}\int_0^{\frac{2\pi}{a}} dk
\left[\sum_{n=1}^N 
\left\langle u_{nk}^{(i)} 
\left| H_k^{(0)} \right|
u_{nk}^{(i)}\right\rangle + 
\left\langle u_{nk}^{(i-1)} \left|
ie\frac{\partial}{\partial k} \right| u_{nk}^{(i)}\right\rangle +
\left\langle u_{nk}^{(i)} \left| ie\frac{\partial}{\partial k} \right|
u_{nk}^{(i-1)}\right\rangle\right.  
\nonumber \\ 
&&\left.- \sum_{m,n=1}^N
\sum_{j,j^\prime=1}^i \sum_{l=0}^{i-1} \delta(2i-j-j^\prime-l)
\Lambda_{mn}^{(l)}(k_j) \left\langle u_{nk}^{(j)} \left|
u_{mk}^{(j^\prime)}\right.\right\rangle\right] \;.
\end{eqnarray}
\begin{eqnarray}
\label{a5}  
E^{(2i+1)} = \frac{a}{\pi} \int_0^{\frac{2\pi}{a}}dk
\left[\sum_{n=1}^N \left\langle u_{nk}^{(i)} \left|
ie\frac{\partial}{\partial k} \right| u_{nk}^{(i)}\right\rangle -
\sum_{n,m=1}^N \sum_{j,j^\prime,l=1}^i \delta(2i+1-j-j^\prime-l)
\Lambda_{mn}^{(l)}(k_j) \left\langle u_{nk}^{(j)} \left|
u_{mk}^{(j^\prime)}\right.\right\rangle\right] \;.\nonumber\\
\end{eqnarray}

Let us examine the $i=1$ and $i=2$ terms in Eq.~\ref{a4}, and the $i=1$
term in Eq.~\ref{a5}.
We use Eq.~\ref{LM1} to write 
\begin{eqnarray}
\label{a6}  
E^{(2)} &=& \frac{a}{\pi}\int_0^{\frac{2\pi}{a}} dk
\sum_{n=1}^N \left\langle u_{nk}^{(1)} \left| 
\left( H_k^{(0)} - \varepsilon_{nk}^{(0)}\right)  \right|
u_{nk}^{(1)}\right\rangle + 
\left\langle u_{nk}^{(0)} \left|
ie\frac{\partial}{\partial k} 
\right| u_{nk}^{(1)}\right\rangle +
\left\langle u_{nk}^{(1)} \left| 
ie\frac{\partial}{\partial k} 
\right| u_{nk}^{(0)}\right\rangle.
\end{eqnarray}

Eq.~\ref{u-ortho} and the orthonormality relation $\left\langle
u_{mk}^{(0)} \left| u_{nk}^{(0)}\right.\right\rangle = \delta_{mn}$
allow us to rewrite the last two terms in this equation in the
explicit gauge-invariant form of Eq.~\ref{E2-invar}, as follows:
\begin{eqnarray}
\label{a7}  
\left\langle \left. u_{nk}^{(1)} \right|\right. 
\frac{\partial}{\partial k} 
\left| \left. u_{nk}^{(0)}\right\rangle \right.
&=& \sum_{m} 
\left\langle \left. u_{nk}^{(1)} \right|\right. 
\partial_k \left. u_{mk}^{(0)}\right\rangle 
\left\langle u_{mk}^{(0)} \left| u_{nk}^{(0)}\right.\right\rangle +
\left\langle u_{nk}^{(1)} \left| u_{mk}^{(0)}\right.\right\rangle 
\left\langle \partial_k
u_{mk}^{(0)} \left| u_{nk}^{(0)}\right\rangle\right.\nonumber\\ 
&=&
\left\langle \left. u_{nk}^{(1)} \right|\right. 
\left( 
\frac{\partial}{\partial k}
\sum_m \left.\left| u_{mk}^{(0)}\right.\right\rangle
\left\langle \left. u_{mk}^{(0)} \right|\right. 
\right) 
\left| \left. u_{nk}^{(0)}\right\rangle\right. \,,
\end{eqnarray}
where we use the notation $\left.\left| \partial_k
u_{mk}\right.\right\rangle \equiv \frac{\partial}{\partial k}
\left.\left| u_{mk} \right.\right\rangle$.  The second term in Eq.~\ref{a6}
is obtained simply as the hermitian conjugate of this latter equation.

For the third and fourth-order terms, we have
\begin{eqnarray}
\label{a8}  
E^{(3)} &=& \frac{a}{\pi}\int_0^{\frac{2\pi}{a}} dk
\sum_{n=1}^N 
\left\langle u_{nk}^{(1)} \left| 
ie\frac{\partial}{\partial k} \right| u_{nk}^{(1)}\right\rangle 
-\sum_{m,n=1}^N 
\left\langle u_{nk}^{(1)} \left| u_{mk}^{(1)}\right.\right\rangle 
\left\langle u_{mk}^{(0)} \left| ie \frac{\partial}{\partial k} \right| 
u_{nk}^{(0)}\right\rangle \,,
\end{eqnarray}
and
\begin{eqnarray}
\label{a9}  
E^{(4)} &=& \frac{a}{\pi}\int_0^{\frac{2\pi}{a}} dk
\sum_{n=1}^N 
\left\langle u_{nk}^{(2)} \left| 
\left( H_k^{(0)} - \varepsilon_{nk}^{(0)}\right)  
\right| u_{nk}^{(2)} \right\rangle + 
\left\langle u_{nk}^{(1)} \left|
ie\frac{\partial}{\partial k} 
\right| u_{nk}^{(2)}\right\rangle +
\left\langle u_{nk}^{(2)} \left| 
ie\frac{\partial}{\partial k} 
\right| u_{nk}^{(1)}\right\rangle  
\nonumber\\
&&-\sum_{m,n=1}^N 
\left(\vphantom{\sum_n^N}
\left\langle u_{nk}^{(1)} \left| u_{mk}^{(2)}\right.\right\rangle 
+\left\langle u_{nk}^{(2)} \left| u_{mk}^{(1)}\right.\right\rangle 
\right)
\left\langle u_{mk}^{(0)} \left| 
ie \frac{\partial}{\partial k} 
\right| u_{nk}^{(0)}\right\rangle\,.
\end{eqnarray}

We now apply Eq.~\ref{a3} to recombine terms in
these two expressions as follows:
\begin{eqnarray}
\label{a10}  
&\mbox{}&\sum_{n=1}^N 
\left\langle u_{nk}^{(1)} \left|
\frac{\partial}{\partial k} 
\right| u_{nk}^{(j)}\right\rangle 
-\sum_{m,n=1}^N 
\left\langle u_{nk}^{(1)} \left| u_{mk}^{(j)}\right.\right\rangle 
\left\langle u_{mk}^{(0)} \left| 
\frac{\partial}{\partial k} 
\right| u_{nk}^{(0)}\right\rangle\nonumber\\ 
&=& \sum_{n=1}^N 
\left\langle \left. u_{nk}^{(1)} \right|\right. 
\left(
\frac{\partial}{\partial k}
\sum_{m=1}^N \left.\left| u_{mk}^{(j)}\right.\right\rangle 
\left\langle \left. u_{mk}^{(0)}\right|\right. 
\right)
\left.\left| u_{nk}^{(0)}\right.\right\rangle\,;
\end{eqnarray}
where $j\!=\!1$ applies to $E^{(3)}$ and $j\!=\!2$ to $E^{(4)}$. This
leads to the gauge-invariant expressions for these quantities,
Eqs.~\ref{E3-invar} and \ref{E4-invar}, respectively.

For the general energy derivative, besides the terms corresponding to
those above for $E^{(3)}$ and $E^{(4)}$, we must also consider terms
of the form
\begin{eqnarray}
\label{a11}  
\Lambda_{mn}^{(l)}(k) \left\langle u_{nk}^{(j)} \left| 
u_{m k}^{(j^\prime)}\right. \right\rangle 
&=& \sum_{l^\prime = 0}^l 
\left\langle u_{mk}^{(l^\prime)} \left| 
H_k^{(0)}
\right| u_{n k}^{(l - l^\prime)} \right\rangle  
\left\langle u_{nk}^{(j)} \left| u_{m k}^{(j^\prime)}\right.\right\rangle 
+  \sum_{l^\prime = 0}^{l-1} 
\left\langle u_{mk}^{(l^\prime)} \left| 
i~e\frac{\partial}{\partial_k}
\right| u_{n k}^{(l-l^\prime-1)} \right\rangle  
\left\langle u_{nk}^{(j)} \left| u_{m k}^{(j^\prime)}\right.\right\rangle \;.\nonumber\\
\end{eqnarray}

For the second term on the right, we use Eqs.~\ref{a1} and \ref{a2} to write 
\begin{eqnarray}
\label{a12}
\left\langle \left. u_{nk}^{(j)} \right|\right. 
\left( 
\sum_{l^\prime = 0}^{l-1} 
\left\langle u_{mk}^{(l^\prime)} \left| \frac{\partial}{\partial_k} 
\right| u_{n k}^{(l-l^\prime-1)} \right\rangle
\right) 
\left.\left| u_{m k}^{(j^\prime)} \right.\right\rangle 
= -\sum_{l^\prime = 0}^{l-1} 
\left\langle \left. u_{nk}^{(j)} \right|\right.  
\left( 
\frac{\partial}{\partial_k} 
\left| \left. u_{mk}^{(j^\prime)}\right.\right\rangle  
\left\langle \left. u_{mk}^{(l^\prime)}\right|\right. 
\right) 
\left.\left| u_{n k}^{(l-l^\prime-1)} \right.\right\rangle\;. 
\end{eqnarray}

These results demonstrate the gauge invariance of the general
expansion terms for the energy. The proper gauge-transformation
properties of the Sternheimer equation can also be proven explicitly along the
same lines, but it follows more simply from the invariance of the
even-order terms for the energy.

\section{}
\label{app:S-series}
In order to develop the PEAD formulation, we examine the following
expression appearing in Eqs.~\ref{P-discr} and \ref{E-def-P}:
\begin{eqnarray}
\label{b1}
ln~det\left[S_{nm}(k_j,k_{j+1}) ({\cal E})\right] &=&
ln~det\left[S^{(0)}_{nm}(k_j,k_{j+1}) + {\cal E}
S^{(1)}_{nm}(k_j,k_{j+1}) + ...\right] \nonumber\\
&=& ln~det\left[S^{(0)}_{nm}(k_j,k_{j+1})\right] + 
\int_0^{\cal E}d{\cal E}\frac{\partial}{\partial{\cal E}}
ln~det\left[S_{nm}(k_j,k_{j+1})({\cal E})\right]\;,
\end{eqnarray}
where the perturbation expansion of $S_{nm}(k_j,k_{j+1}) ({\cal E})$ is defined
according to Eq.~\ref{series}.  With the help of the ``magic'' formula
\begin{eqnarray}
\label{b2}
\frac{\partial}{\partial
\lambda}\left\{{\vphantom{\frac{a}{b}}}ln~det\left[A(\lambda)\right]\right\}
= tr\left[A^{-1}\frac{\partial A}{\partial\lambda}\right] =
\sum_{m,n}A^{-1}_{mn}\frac{\partial A_{nm}}{\partial\lambda}\,,
\end{eqnarray}
we can rewrite Eq.~\ref{b1} (to simplify the notation, in the
following equations we drop the $k$-point arguments) as follows:
\begin{eqnarray}
\label{b3}
ln~det\left[S_{nm} ({\cal E})\right] =
ln~det\left[S^{(0)}_{nm}\right] + \int_0^{\cal E}d{\cal E}
\sum_{mn}\frac{\partial S_{nm} ({\cal E})}{\partial{\cal E}}
S^{-1}_{mn}({\cal E})\;.
\end{eqnarray}

To proceed further, we let $\Delta S({\cal E}) = S({\cal E}) - S^{(0)}
=  {\cal E} S^{(1)}({\cal E}) + {\cal E}^2 S^{(2)}({\cal E}) + {\cal E}^3
S^{(3)}({\cal E}) + ...$, and the inverse overlap matrix can be written
as
\begin{eqnarray}
\label{b4}
S^{-1}({\cal E}) = \left\{ S^{(0)}\left[ 
I + Q\Delta S({\cal E})\right]\right\}^{-1} = 
\left\{ I + \sum_{i=1}^{\infty}(-1)^i\left[{\vphantom \sum}Q
\Delta S({\cal E})\right]^i\right\}Q\;,
\end{eqnarray}
where, following the notation introduced in Sec.~\ref{lower-P}, we
use $Q = [S^{(0)}]^{-1}$ for the inverse of the zeroth-order overlap
matrix.

To obtain the lower-order terms in the expansion of Eq.~\ref{b1}, we
write the terms up to third-order explicitly:
\begin{eqnarray}
\label{b5}
S^{-1}({\cal E}) &=& Q - {\cal E} Q S^{(1)} Q
+  {\cal E}^2\left[ - Q S^{(2)} Q +  
Q S^{(1)} Q S^{(1)} Q \right] \nonumber \\
&+& 
{\cal E}^3\left[ - Q S^{(3)} Q +  
Q S^{(2)} Q S^{(1)} Q +
Q S^{(1)} Q S^{(2)} Q \right.\nonumber \\
&-&\left. Q S^{(1)} Q S^{(1)} Q S^{(1)} Q \right]
+ {\cal O}({\cal E}^4)\;.
\end{eqnarray}

Combining Eqs.~\ref{b3}, \ref{b5} and $\partial S({\cal E})/
\partial{\cal E} = S^{(1)} + 2{\cal E}S^{(2)} + 3{\cal E}^2S^{(3)} +
4{\cal E}^3S^{(4)} + {\cal O}({\cal E}^4)$, we arrive at
\begin{eqnarray}
\label{b6}
ln~det\left[S({\cal E})\right] &=& ln~det\left[S^{(0)}\right] 
+ {\cal E} tr\left[ S^{(1)}Q\right] +  
  \frac{{\cal E}^2}{2} {\rm Tr} \left[ 2 S^{(2)} Q - 
S^{(1)} Q S^{(1)} Q \right] \nonumber \\
&+& 
\frac{{\cal E}^3}{3} tr\left[ 3 S^{(3)} Q -  
3 S^{(2)} Q S^{(1)} Q +
S^{(1)} Q S^{(1)} Q S^{(1)} Q \right] \nonumber \\
&+&\frac{{\cal E}^4}{4} tr\left[ 4 S^{(4)} Q 
-  4 S^{(3)} Q S^{(1)} Q 
-2 S^{(2)} Q S^{(2)} Q \right.\nonumber \\ 
&+&\left. 4 S^{(2)} Q S^{(1)} Q S^{(1)} Q  
-S^{(1)} Q S^{(1)} Q S^{(1)} Q S^{(1)} Q 
\right] + {\cal O}({\cal E}^5)\;.
\end{eqnarray}

Finally, we note that the magic formula may also be used to derive the
Euler-Lagrange equation given in Eq.~\ref{stern0-P}, as follows:
\begin{eqnarray}
\frac{\delta\left\{\ln\det\!~\!\left[\frac{\vphantom{}}{\vphantom{}}
S_{nm}(k_j,k_{j+1}) \right]\right\}} {\delta u_{nk}^\ast}
= {\rm Tr}\left[ S^{-1}\frac{\delta S}{\delta u_{nk}^\ast}\right]
= \sum_{m=1}^N \left.\left| u_{mk_{j+1}} \right.\right\rangle
S^{-1}_{mn}(k_j,k_{j+1}) \;.
\end{eqnarray}
%

%
%



\section{}
\label{app:1dmodel}
In this appendix, we provide all the steps for the derivation of the
energy derivatives for the 1D-TB model, given in
Sec.~\ref{sec:1d-model}. The expressions for the discretized versions of
$E^{(4)}$ are not written explicitly.

Eq.~\ref{1d-E0} for the unperturbed total energy is a
complete elliptic integral of the second kind~\cite{mathfis}, which is given in
its general form by
\begin{eqnarray}
\label{ellipt}
{\cal I}_{n \over 2} = \int_0^{\pi \over 2} dy \left( 1 + A^2 \cos^2
y \right)^{n \over 2}\;,
\end{eqnarray}
where $n$ is a positive or negative integer, and $A = 4 t$. Several such
integrals will be encountered in the course of our derivation.

 
\subsection{Continuous formulation}
\label{app:1dmodel.1}
We begin by developing some useful preliminary results. For our 1D-TB
model, the derivatives with respect to $k$ of the unperturbed valence
and conduction states (Eq.~\ref{1d-eigstate}) are written analytically
as
\begin{eqnarray}
\label{dudk}
i\frac{\partial}{\partial k} 
\left.\left| u_{vk}^{(0)} \right.\right\rangle = - 
\frac{\partial \alpha_{vk}}{\partial k}  
\left.\left| u_{vk}^{(0)} \right.\right\rangle 
-i \frac{\partial \Theta_k}{\partial k} e^{-i\Delta\alpha_k} 
 \left.\left| u_{ck}^{(0)} \right.\right\rangle \nonumber \\
i\frac{\partial}{\partial k}  
\left.\left| u_{ck}^{(0)} \right.\right\rangle = - 
\frac{\partial \alpha_{ck}}{\partial k}  
\left.\left| u_{ck}^{(0)} \right.\right\rangle 
+i \frac{\partial \Theta_k}{\partial k} e^{i\Delta\alpha_k}
 \left.\left| u_{vk}^{(0)} \right.\right\rangle \;,
\end{eqnarray}
where $\Delta\alpha_k =\alpha_{ck}-\alpha_{vk}$. 

Recalling that $\left\langle u_{vk}^{(0)} \left| u_{vk}^{(1)}
\right.\right\rangle = 0 $, hence \mbox{$P_{vk} \left.\left|
u_{vk}^{(1)} \right.\right\rangle = 0$}, the solution of
Eq.~\ref{1d-str1} gives the first-order change of the valence state
\begin{eqnarray}
\label{1d-v1}
 \left.\left| u_{vk}^{(1)}\right.\right\rangle = 
P_{ck} \left.\left| u_{vk}^{(1)} \right.\right\rangle = {i
\over \Delta\varepsilon_k} {\partial \Theta_k \over \partial k}
e^{-i\Delta\alpha_k}\left.\left| u_{ck}^{(0)} \right.\right\rangle \;,
\end{eqnarray}
where $\Delta\varepsilon_k = \varepsilon_{ck}^{(0)} -
\varepsilon_{vk}^{(0)} = \left[ 1 + A^2 \cos^2 \left( {k \over 2}
\right) \right]^2$.  From Eqs.~\ref{1d-eigval} and \ref{theta}, the
partial derivative appearing in this expression is simply
\begin{eqnarray}
\label{dthetadk}
{\partial \Theta_k \over \partial k} = \frac{t~\!sin \left( {k \over 2}
\right)}{\left[ 1 + A^2 \cos^2 \left( {k \over 2} \right) \right]
}\;.
\end{eqnarray}

Using Eq.~\ref{dthetadk}, the second-order energy term
(Eq.~\ref{1d-E2}) is written
\begin{eqnarray}
\label{1d-E2.a}
E^{(2)} = {1 \over 4 \pi} \left[ {\cal I}_{-{3 \over 2}} - \left(1 +
A^2\right) {\cal I}_{-{5 \over 2}} \right]\;.
\end{eqnarray}

The second-order change of the valence state is obtained from
Eqs.~\ref{stern2-invar} and \ref{Pv-u2}. The first-order Lagrange
multiplier in Eq.~\ref{stern2-invar} is obtained from
Eqs.~\ref{1d-eigstate}, \ref{LM1}, and \ref{dudk}:
\begin{eqnarray}
\label{1d-lagr1}
\Lambda_k^{(1)} = \left\langle u_{vk}^{(0)} \left| 
i\frac{\partial}{\partial k} \right| u_{vk}^{(0)} \right\rangle 
= - \frac{\partial \alpha_{vk}}{\partial k}\;.
\end{eqnarray}
Simple algebraic manipulations involving Eqs.~\ref{dudk},~\ref{1d-v1},
and \ref{1d-lagr1}, combined with Eqs.~\ref{stern2-invar} and
\ref{Pv-u2}, yield the second-order wave-function derivatives
\begin{eqnarray}
\label{1d-u2}
P_{ck} \left.\left| u_{vk}^{(2)} \right.\right\rangle &=& {1 \over \Delta\varepsilon_k } 
\left[ {\partial \over
\partial k} \left( {1 \over \Delta\varepsilon_k} 
{\partial \Theta_k \over \partial k} \right)\right] 
e^{-i \Delta \alpha_k}\left. \left| u_{ck}^{(0)}\right.\right\rangle\,,
\nonumber \\
P_{vk} \left.\left| u_{vk}^{(2)} \right.\right\rangle &=& -{1 \over 2} 
 \left.\left| u_{vk}^{(0)} \right.\right\rangle 
\left\langle u_{vk}^{(1)} \left| u_{vk}^{(1)} \right.\right\rangle = 
-{1 \over 2} \left( {1 \over \Delta\varepsilon_k} 
{\partial \Theta_k \over \partial k} \right)^2 
\left.\left| u_{vk}^{(0)} \right.\right\rangle \;.
\end{eqnarray}

These results lead to Eq.~\ref{1d-E4} for
$E^{(4)}$.  In terms of elliptic
integrals, this quantity is written
\begin{eqnarray}
\label{1d-E4.a}
E^{(4)} &=& \frac{37\left(1 + A^2\right)^2}{64\pi} {\cal I}_{-{11 \over 2}} 
- \frac{\left(1 + A^2 \right)}{32\pi} \left[ 18 \left( 1 + A^2 \right) + 25 \right] 
{\cal I}_{-{9 \over2}} 
+ \frac{1}{64\pi} \left[ 48 \left(
1 + A^2 \right) + 17 \right] {\cal I}_{-{7 \over 2}} \nonumber \\
&& - \frac{1}{4\pi}{\cal I}_{-{5
\over 2}}\;.
\end{eqnarray}  
%


\subsection{DAPE formulation}

We now apply the DAPE expressions obtained in Sec.~\ref{discr}
to our 1D-TB model. The first-order valence state is given by
\begin{eqnarray}
\label{1d-str1-discr}
&& P_{ck_j} \left[ H_{k_j}^{(0)} - \varepsilon_{vk_j}^{(0)} \right]
P_{ck_j} \left.\left| u_{vk_j}^{(1)} \right.\right\rangle = 
- {i \over 2\Delta k} P_{ck_j} 
\left[ \vphantom{\sum_n^N} \left.\left| u_{v k_{j+1}}^{(0)} \right.\right\rangle \left\langle
u_{v k_{j+1}}^{(0)} \left| u_{vk_j}^{(0)} \right.\right\rangle - \left.\left| u_{v k_{j-1}
}^{(0)} \right.\right\rangle \left\langle u_{v k_{j-1} }^{(0)} \left| u_{vk_j}^{(0)}
\right.\right\rangle \right] 
\nonumber\\ 
&& \; = - {i \over 2\Delta k} \left.\left| u_{ck_j}^{(0)} \right.\right\rangle 
\left[ \vphantom{\sum_n^N} \left\langle u_{ck_j}^{(0)} 
\left| u_{v k_{j+1}}^{(0)} \right.\right\rangle \left\langle u_{vk_{j+1}}^{(0)} 
\left| u_{vk_j}^{(0)} \right.\right\rangle - \left\langle u_{c k_j}^{(0)}
\left| u_{v k_{j-1} }^{(0)} \right.\right\rangle \left\langle u_{v k_{j-1} }^{(0)} 
\left| u_{vk_j}^{(0)} \right.\right\rangle \right]\;.
\end{eqnarray}

Using Eq.~\ref{1d-eigstate}, this expression becomes
\begin{eqnarray}
\label{v1-discr}
\left.\left| u_{vk_j}^{(1)} \right.\right\rangle = {i \over 2 \Delta k \Delta\varepsilon_j } 
\sin\left(\Theta_{j+1} -  \Theta_{j-1}\right)
\cos\left(2\Theta_j - \Theta_{j+1} -  \Theta_{j-1}\right) 
e^{-i \Delta \alpha_j} \left.\left| u_{ck_j}^{(0)} \right.\right\rangle\;;
\end{eqnarray}
where we use a simplified notation $\Theta_{k_j}\rightarrow \Theta_j$
(likewise for $\Delta \alpha$ and $\Delta \varepsilon$). From 
Eqs.~\ref{v1-discr} and \ref{E2-discr-nv}, we obtain
Eq.~\ref{1d-E2-discr}.

The second-order wavefunction is obtained from 
Eqs.~\ref{Pv-u2} and \ref{stern2-discr}. The Sternheimer
equation is written
\begin{eqnarray}
\label{1d-str2-discr}
&& P_{ck_j} \left[ H_{k_j}^{(0)} - \varepsilon_{vk_j}^{(0)} \right] P_{ck_j} 
\left.\left| u_{vk_j}^{(2)} \right.\right\rangle = - {i \over 2\Delta k}P_{ck_j}   
\left[\vphantom{\sum_n^N} \left.\left| u_{v k_{j+1}}^{(1)} \right.\right\rangle 
\left\langle u_{v k_{j+1}}^{(0)} \left| u_{vk_j}^{(0)} \right.\right\rangle -
\left.\left| u_{v k_{j-1} }^{(1)} \right.\right\rangle 
\left\langle u_{v k_{j-1} }^{(0)} \left| u_{vk_j}^{(0)} \right.\right\rangle 
\right] \nonumber \\
&&\; = - {i \over 2\Delta k} \left.\left| u_{ck_j}^{(0)} \right.\right\rangle
\left[\vphantom{\sum_n^N} \left\langle u_{c k_j}^{(0)} 
\left| u_{v k_{j+1}}^{(1)} \right.\right\rangle 
\left\langle u_{v k_{j+1}}^{(0)} \left| u_{vk_j}^{(0)} \right.\right\rangle -
\left\langle u_{c k_j}^{(0)} \left| u_{v k_{j-1} }^{(1)} \right.\right\rangle 
\left\langle u_{v k_{j-1} }^{(0)} \left| u_{vk_j}^{(0)} \right.\right\rangle 
\right]\;.
\end{eqnarray}

By combining Eqs.~\ref{1d-eigstate}, \ref{Pv-u2}, and \ref{v1-discr}, we arrive at
\begin{eqnarray}
\label{v2-discr}
P_{vk_j} \left.\left| u_{vk_j}^{(2)} \right.\right\rangle &=& 
- {1 \over 8 \Delta k^2 \Delta\varepsilon_j^2} 
\sin^2 \left(\Theta_{j+1} - \Theta_{j-1}\right) 
\cos^2 \left(2\Theta_j -\Theta_{j+1} - \Theta_{j-1} \right)
\left.\left| u_{vk_j}^{(0)}\right.\right\rangle \nonumber \\
P_{ck_j} \left.\left| u_{vk_j}^{(2)} \right.\right\rangle &=&
\frac{e^{-i \Delta \alpha_j}}
{4 \Delta k^2 \Delta\varepsilon_j}\left[
{1 \over \Delta\varepsilon_{j+1}}
\sin\left(\Theta_{j+2} - \Theta_j \right)
\cos\left( 2\Theta_{j+1} -\Theta_{j+2} -\Theta_j \right)
\cos^2\left( \Theta_{j+1}-\Theta_j \right) \right. \nonumber \\
&&-{1 \over \Delta\varepsilon_{j-1}}\left.
\sin\left(\Theta_j -\Theta_{j-2}\right) 
\cos\left( 2\Theta_{j-1} -\Theta_{j-2} -\Theta_j \right)
\cos^2\left( \Theta_j - \Theta_{j-1} \right)
\vphantom{\sum_n^N}\right] \left.\left| u_{ck_j}^{(0)} \right.\right\rangle \,.
\end{eqnarray}

Eqs.~\ref{v1-discr} and \ref{v2-discr}, combined with
Eq.~\ref{E4-discr}, lead to an analytic expression for $E^{(4)}$.


\subsection{PEAD formulation}

Here, we apply the PEAD expressions from Sec.~\ref{sec:pt-P} to the
model system. The zero-field overlap matrix, \mbox{$S^{(0)}\left( k_j,
k_{j+1} \right) = \left\langle u_{vk_j}^{(0)} \left| u_{vk_{j+1}}^{(0)}
\right.\right\rangle$} is 
\begin{eqnarray}
\label{1d-S0}
S^{(0)}\left( k_j, k_{j+1} \right) =
e^{i\left(\alpha_{v_{j+1}}-\alpha_{v_j}\right)}
\cos\left(\Theta_{j+1}-\Theta_j \right)\;.  
\end{eqnarray}
Note that for this model, $S$ is a 1$\times $1 matrix, and hence
the inverse overlap matrix is simply 

\begin{eqnarray}
\label{1d-Q}
Q\left( k_j, k_{j+1} \right) = \frac{1}{S^{(0)}\left( k_j, k_{j+1} \right)}
=\frac {e^{ - i\left(\alpha_{v_{j+1}}-\alpha_{v_j}\right)}}
{\cos\left(\Theta_{j+1}-\Theta_j \right)}\;.
\end{eqnarray}

From Eq.~\ref{stern1-P}, the first-order Sternheimer equation
reads
\begin{eqnarray}
\label{P1d-str1}
P_{ck_j} \left[ H_{k_j}^{(0)} - \varepsilon_{vk_j}^{(0)} \right] 
P_{ck_j} \left.\left| u_{vk_j}^{(1)} \right.\right\rangle = 
- P_{ck_j} {i \over 2\Delta k} \left[ \vphantom{\sum_n^N}
\left.\left| u_{v k_{j+1}}^{(0)} \right.\right\rangle Q\left( k, k_{j+1} \right) - 
\left.\left| u_{v k_{j-1} }^{(0)} \right.\right\rangle Q\left( k, k_{j-1} \right)
\right]\;.
\end{eqnarray}
Using Eq.~\ref{1d-eigstate}, we obtain
\begin{eqnarray}
\label{P1d-v1}
\left.\left| u_{vk_j}^{(1)} \right.\right\rangle = 
\frac{i e^{- i \Delta \alpha_j}} {2 \Delta k \Delta\varepsilon_k} 
\left.\left| u_{ck}^{(0)} \right.\right\rangle
\left[\vphantom{\sum_n^N} \tan\left(\Theta_{j+1} -  \Theta_j \right)
-\tan\left(\Theta_{j-1} - \Theta_j \right)\right] \;.
\end{eqnarray}

To obtain the second-order wavefunctions, we need the first-order term
for the overlap matrix $S^{(1)}(k_j, k_{j+1}) = \left\langle u_{v
k_j}^{(0)} \left| u_{v k_{j+1}}^{(1)} \right.\right\rangle + \left\langle u_{v k_j}^{(1)}
\left| u_{v k_{j+1}}^{(0)} \right.\right\rangle$. Thanks to Eqs.~\ref{1d-eigstate}
and \ref{P1d-v1}, we have
\begin{eqnarray}
\label{1d-S1}
S^{(1)}\left( k_j, k_{j+1} \right) &=& \frac{i e^{i 
\left(\alpha_{v_{j+1}} - \alpha_{j}\right)}}{2\Delta k} 
\sin \left(\Theta_{j+1} - \Theta_j\right)
\left[\frac{\tan\left(\Theta_{j+2} -\Theta_{j+1}\right) - 
\tan\left(\Theta_{j} -\Theta_{j+1}\right)}{\Delta\varepsilon_{j+1}}~+
\right. \nonumber\\
&&\left.\frac{\tan\left(\Theta_{j+1} -\Theta_{j}\right) - 
\tan\left(\Theta_{j-1} -\Theta_{j}\right)}{\Delta\varepsilon_{j}}\right]\,.
\end{eqnarray}

Plugging the above results for $Q$, $S^{0}$, $S^{(1)}$,
$u_{vk}^{(0)}$, and $u_{vk}^{(1)}$ in Eqs.~\ref{Pv-u2} and \ref{stern2-P} we get
\begin{eqnarray}
\label{P1d-v2}
P_{vk_j} \left.\left| u_{vk_j}^{(2)} \right.\right\rangle &=& 
- {1 \over 8 \Delta k^2 \Delta\varepsilon_j^2} 
\left.\left| u_{vk}^{(0)}\right.\right\rangle \left[\tan\left(\Theta_{j+1}-\Theta_j\right)
-\tan\left(\Theta_{j-1}-\Theta_j\right)\right]^2 \nonumber \\
P_{ck_j} \left.\left| u_{vk_j}^{(2)} \right.\right\rangle &=&
\left.\left| u_{ck_j}^{(0)} \right.\right\rangle \frac{e^{-i \Delta \alpha_j}}
{4\Delta^2\!k \Delta\varepsilon_j}\left\{
\frac{\left[1+\tan^2\left(\Theta_{j+1} - \Theta_j \right)\right]
\left[\tan\left(\Theta_{j+2} - \Theta_{j+1} \right) -
\tan\left(\Theta_{j} - \Theta_{j+1} \right)\right]}{\Delta\varepsilon_{j+1}}
\right. +\nonumber\\
&&\frac{\left[1+\tan^2\left(\Theta_{j-1} - \Theta_j \right)\right]
\left[\tan\left(\Theta_{j-2} - \Theta_{j-1} \right) -
\tan\left(\Theta_{j} - \Theta_{j-1} \right)\right]}{\Delta\varepsilon_{j-1}}
+\nonumber\\
&&\left.\frac{\left[\tan^2\left(\Theta_{j+1} - \Theta_j \right) - 
\tan^2\left(\Theta_{j-1} - \Theta_j \right)\right]
\left[\tan\left(\Theta_{j+1} - \Theta_{j} \right) -
\tan\left(\Theta_{j-1} - \Theta_{j-1} \right)\right]}{\Delta\varepsilon_{j}}
\right\}\;.
\end{eqnarray}

As in the DAPE case, from these expressions for $u^{(1)}_{vk_j}$ and
$u^{(2)}_{vk_j}$ follows the PEAD analytic form for $E^{(4)}$.

\begin{figure}[t]
\centering{
\vspace{0.25cm}
\caption{The phase $\gamma_k$ in the trial wavefunctions for the 1D
model.  $\gamma_k$ changes by $2\pi$ over a small interval $\Delta k$
centered at an arbitrary $k$-vector in the Brillouin zone. The
position of the jump, $\langle k \rangle$, is indicated in the figure by
the vertical dotted line.}
\label{gamma-k}
}
\end{figure}
\begin{figure}[t]
\centering{
\vspace{0.25cm}
\caption{The function $\cos \gamma_k$, which differs from 1 over an interval 
$\Delta k$ in the Brillouin zone.}
\label{cos-gamma}
}
\end{figure}
\begin{figure}[t]
\centering{
\vspace{0.25cm}
\caption{The percentual error $\left[ E^{(2)}_{discr.} -
E^{(2)}_{exact}\right]/E^{(2)}_{exact}$ in the evaluation of the
second-order change in energy for 1D model using the DAPE and the PEAD
formulations, with a 20 $k$-point sampling of the Brillouin zone. The
hoping parameter varies over the [0,1] interval.}
\label{e2fig}
}
\end{figure}
\begin{figure}[t]
\centering{
\vspace{0.25cm}
\caption{The percentual error $\left[ E^{(4)}_{discr.} -
E^{(4)}_{exact}\right]/E^{(4)}_{exact}$ in the evaluation of the
second-order change in energy for 1D model using the DAPE and the PEAD
formulations, with an 80 $k$-point sampling of the Brillouin zone. The
hoping parameter varies over the [0,1] interval.}
\label{e4fig}
}
\end{figure}


\begin{references}
\bibitem{footnote3} Local-field effects are associated with the Umklapp
terms in the dielectric response: a macroscopic electric field
induces modifications of the density at the atomic scale, that induce
modifications of the potential at the atomic scale, that induce
modifications of the macroscopic polarization.
In a more formal approach, they arise
because the macroscopic
dielectric constant is the inverse of the
``head'' (G=0,G'=0) of the inverse
dielectric matrix (see Ref. ~\onlinecite{hybers}),
and not simply the ``head'' of the dielectric matrix.
\bibitem{footnote-gauge} Eq.~\ref{gauge-invar} expresses the fact that
these quantities are actually the perturbation expansion of the
projector onto the occupied space of the problem, thus being gauge
invariant by definition.
\end{references}
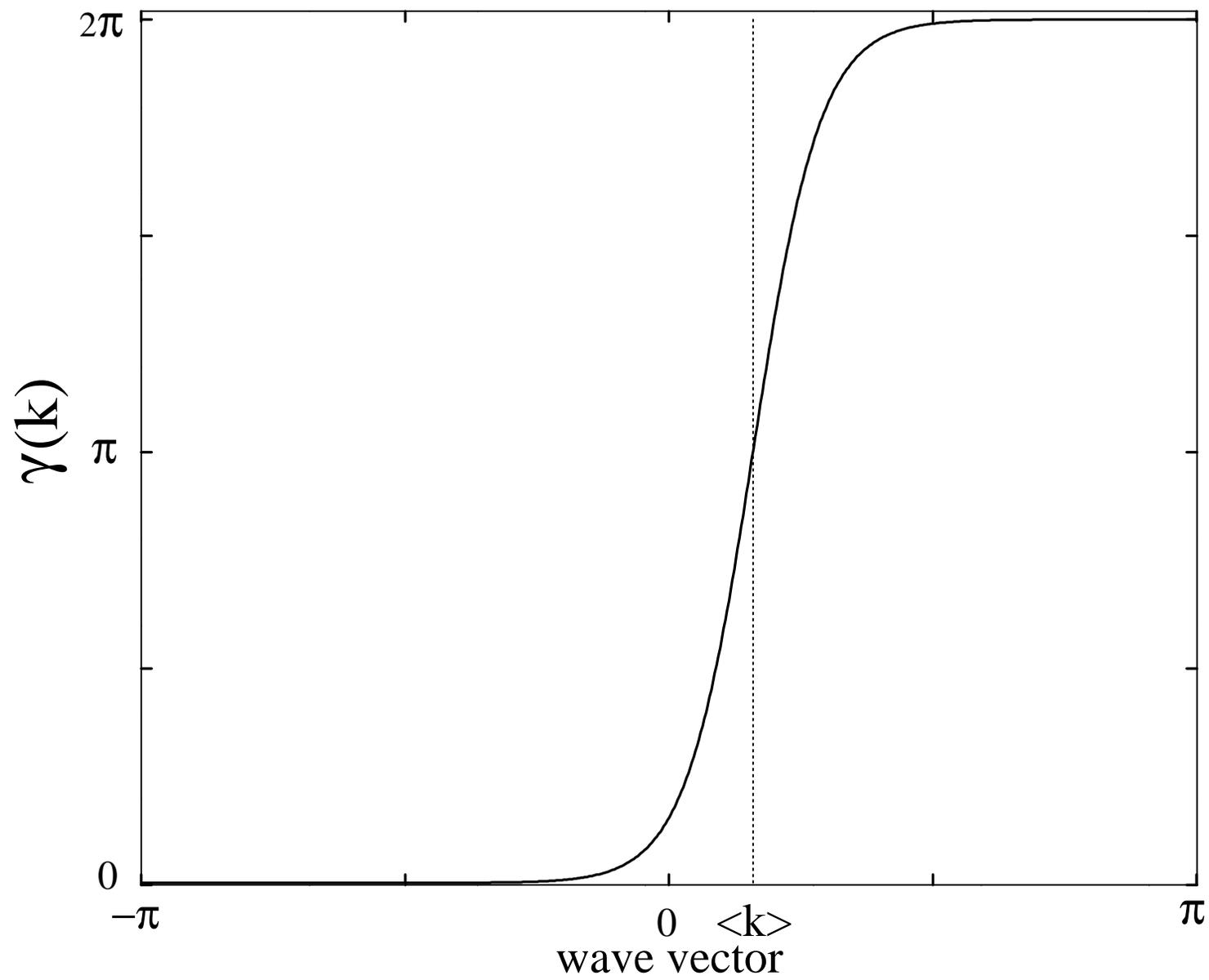
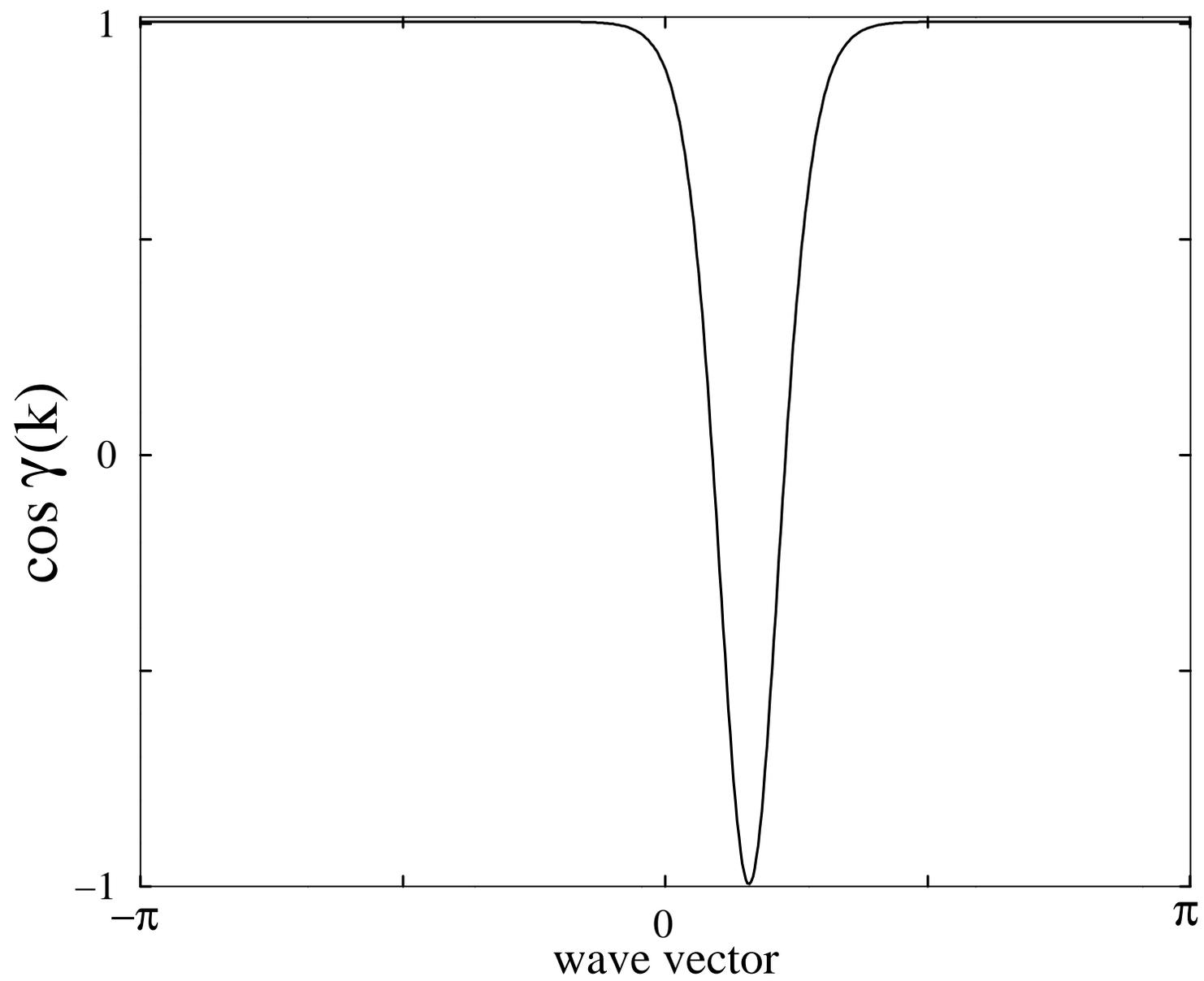
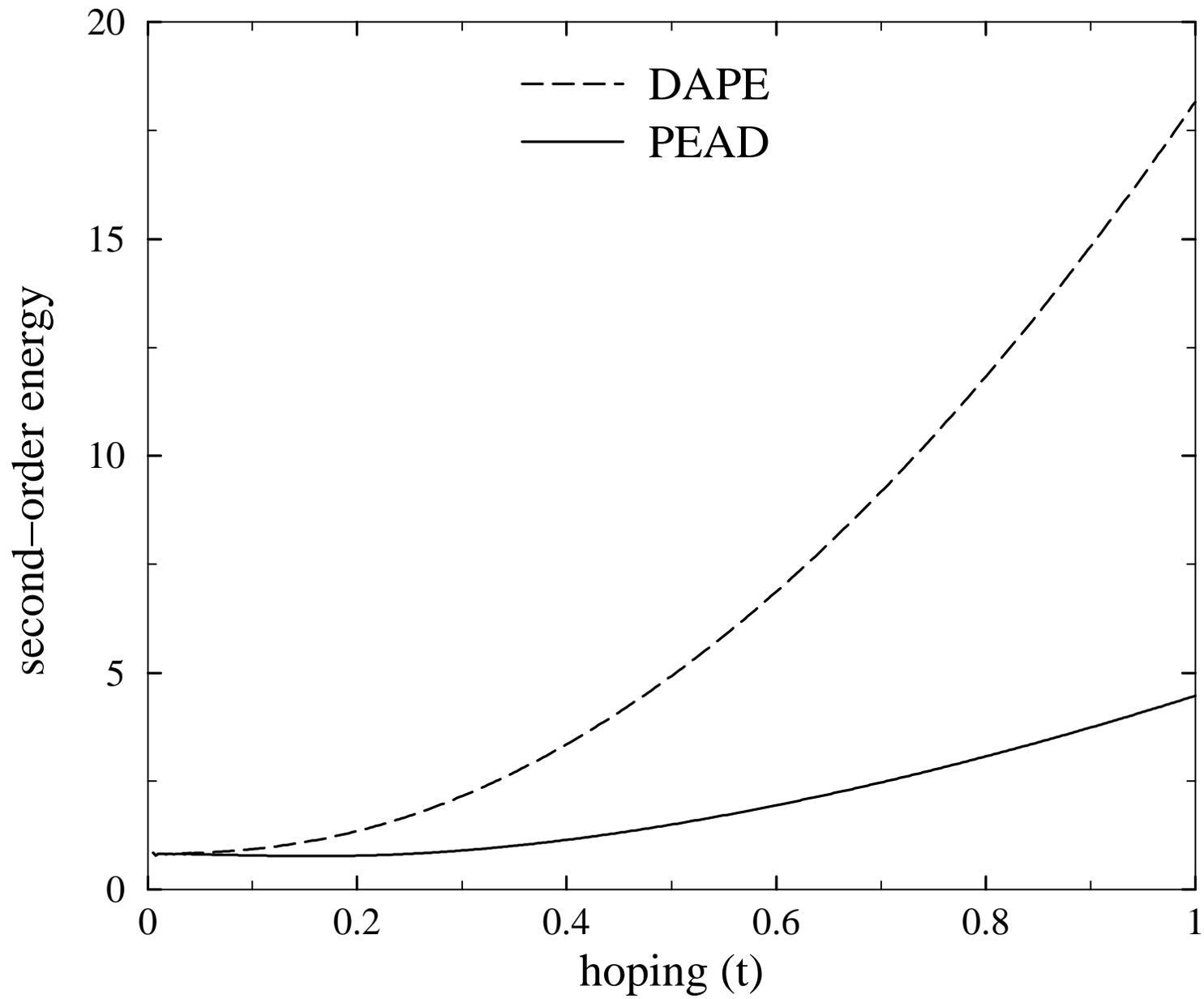
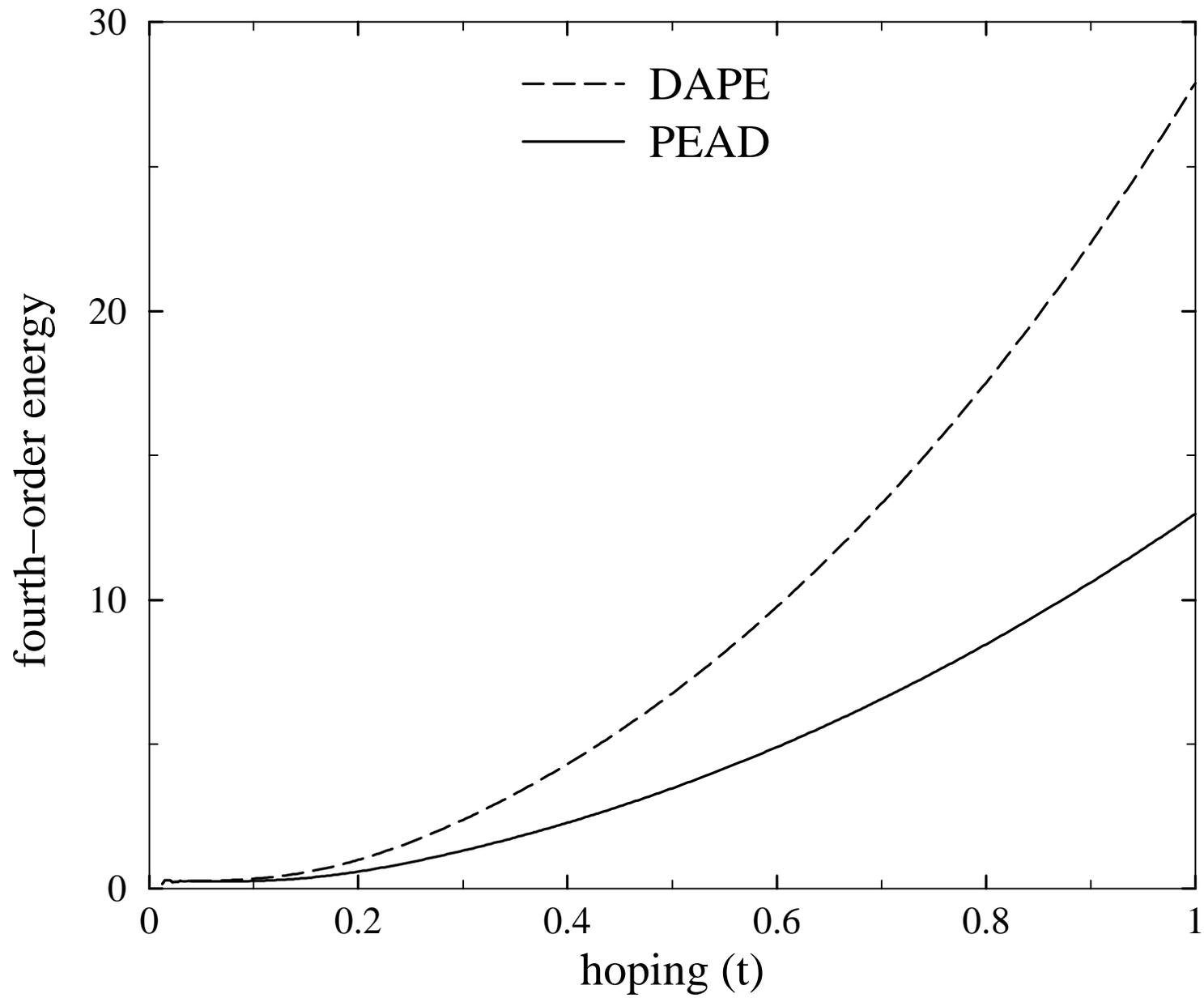
\end{document}